# The Linear-Time-Invariance Notion of the Koopman Analysis —Part 2: Physical Interpretations of Invariant Koopman Modes and Phenomenological Revelations

Cruz Y. Li[12] (李雨桐), Zengshun Chen[1*] (陈增顺), Tim K.T. Tse[3**] (谢锦添), Asiri Umenga Weerasuriya[4], Xuelin Zhang[5] (张雪琳), Yunfei Fu[6] (付云飞), Xisheng Lin[7] (蔺习升)

[1] *Department of Civil Engineering, Chongqing University, Chongqing, China*

[2,3,4,6,7] *Department of Civil and Environmental Engineering, The Hong Kong University of Science and Technology, Hong Kong SAR, China*

[5] *School of Atmospheric Sciences, Sun Yat-sen University, Zhuhai, China.*

[1] yliht@connect.ust.hk; ORCID 0000-0002-9527-4674

[2] zchenba@connect.ust.hk; ORCID 0000-0001-5916-1165

[3] timkttse@ust.hk; ORCID 0000-0002-9678-1037

[4] asiriuw@connect.ust.hk; ORCID 0000-0001-8543-5449

[5] zhangxlin25@mail.sysu.edu.cn; ORCID 0000-0003-3941-4596

[6] yfuar@connect.ust.hk; ORCID 0000-0003-4225-081X

[7] xlinbl@connect.ust.hk; ORCID 0000-0002-1644-8796

[*] Co-first author with equal contribution.

[**] Corresponding author

All correspondence is directed to Dr. Tim K.T. Tse.

# Abstract

This serial work presents a Linear-Time-Invariance (LTI) notion to the Koopman analysis, finding consistent and physically meaningful Koopman modes and addressing a long-standing problem of fluid-structure interactions: deterministically relating the fluid and structure. Part 1 (*Li et al.*, 2022) developed the Koopman-LTI architecture and applied it to a pedagogical prism wake. By the systematic procedure, the LTI generated a sampling-independent Koopman linearization that captured all the recurring dynamics, finding six corresponding, orthogonal, and in-synch fluid excitation-structure response mechanisms. This Part 2 analyzes the six modal duplets' to underpin their physical interpretations, providing a phenomenological revelation of the subcritical prism wake. By the dynamical mode shape, results show that two mechanisms at $St_1=0.1242$ and $St_5=0.0497$ describe shear layer dynamics, the associated Bérnard-Kármán shedding, and turbulence production, which together overwhelm the upstream and crosswind walls by instigating a reattachment-type of response. The on-wind walls' dynamical similarity renders them a spectrally unified fluid-structure interface. Another four harmonic counterparts, namely the subharmonic at $St_7=0.0683$, the second harmonic at $St_3=0.2422$, and two ultra-harmonics at $St_7 =0.1739$ and $St_{13}=0.1935$, govern the downstream wall. The 2P wake mode is also observed as an embedded harmonic of the bluff-body wake. Finally, this work discovered the vortex breathing phenomenon, describing the constant energy exchange in wake's circulation-entrainment-deposition processes. With the Koopman-LTI, one may pinpoint the exact excitations responsible for a specific structural response, or vice versa.

# 1. Introduction

The omnipresence of fluid-structure interactions (FSI) demands persistent research and exploration. To date, fluids' volatility, their nonlinear interactive mechanisms with structures, and the unsolved Navier-Stokes equations leave FSI a persisting enigma of fluid mechanics. Fortunately, recent advancements in data science offered a brand-new pathway to the solution (Budišić et al., 2012; Kutz et al., 2016; Lusch et al., 2018; Raissi et al., 2019). Seven decades after the Koopman theory (Koopman, 1931; Koopman & Neumann, 1932), its mathematical promise was brought to life by the pioneers of the data-driven approach (Mauroy & Mezić, 2012; Mezić, 2005; Rowley et al., 2009; Schmid, 2010). Since then, the Koopman analysis has been pervasively applied to fluid problems with success (Carlsson et al., 2014; Dotto et al., 2021; Eivazi et al., 2020; Garicano-Mena et al., 2019; Herrmann et al., 2021; Jang et al., 2021; N.-H. Liu et al., 2021; Q. Liu et al., 2021; Y. Liu et al., 2021; Magionesi et al., 2018; Miyanawala & Jaiman, 2019; Muld et al., 2012; Page & Kerswell, 2019; W. Wu et al., 2020).

On this celebratory note, we observed that the most organized research efforts were led by applied mathematicians and focused on the algorithmic development of the Koopman analysis (Brunton et al., 2016; Budišić et al., 2012; Mezić, 2013; Rowley & Dawson, 2017; Schmid et al., 2011; Taira et al., 2017). The analytical end is left to individual interpretations; hence it is usually case-specific and incohesive. This serial effort focuses on organizing and developing the analytical end, specifically on a long-standing issue of FSI: deterministically finding and relating an observed structure response to its flow field origin. In Part 1, we proposed the linear-time-invariance (LTI) notion and developed an analytical procedure, called the Koopman-LTI architecture (see figure 1), to systematically address FSI problems. Through a pedagogical demonstration on a prism wake with the most rudimentary Dynamic Mode Decomposition algorithm, Part 1 successfully:

1. Generated a sampling-independent Koopman linearization that captured all the prominent recurring dynamics. The mean reconstruction error, rms reconstruction error, and DMD approximation error of the Koopman modes were all numerical zeros, $O^{-12}$, $O^{-9}$, and $O^{-8}$, respectively.
2. Revealed *w*'s trivial role in the convection-dominated free-shear flow, Reynolds stresses' spectral description of cascading eddies, vortices' sensitivity to dilation and indifference to distortion, and structure responses' origin in vortex activities.
3. Reduced the subcritical wake during shear layer transition II into only six dominant excitation-response dynamics. The upstream and crosswind walls constitute a dynamically unified interface, which is dominated by only two mechanisms at $St=0.1242$ and $St=0.0497$ (Class 1). The downstream wall remains a distinct interface and is dominated by four other mechanisms at $St=0.1739$, $St=0.0683$, $St=0.1925$, and $St=0.2422$ (Class 2).

By completing the *Input Curation*, *Koopman Algorithm*, *Linear-Time-Invariance*, and *Constitutive Relationship* modules, the fluid-structure constitution has been established. The complete revelation essentially comes down to understanding the six mechanisms, which this Part 2 will address through the final *Phenomenological Relationship* module.

Phenomenology, the study of phenomena, is the essential path to solution for our test subject. As Roshko (1993) remarked:

> *"The problem of bluff body flow remains almost entirely in the empirical, descriptive realm of knowledge."*

In this paper, we visualize the Koopman modes to give phenomenological consolidation to the abstract fluid-structure constitution. The visualization is modal-based, therefore, normalizable for inter-mode comparison. We also propose a methodical improvement, the dynamic mode shape, for visualizing coherent structures (Hussain, 1986). It includes phase information and offers in-synch portrayals of the structure and flow field's behaviors.

In composition, Section 2 analyzes the *Class 1* mechanisms. Section 3 focuses on the *Class 2* mechanisms. Section 4 presents a newly discovered vortex breathing phenomenon. Section 5 offers a summary.

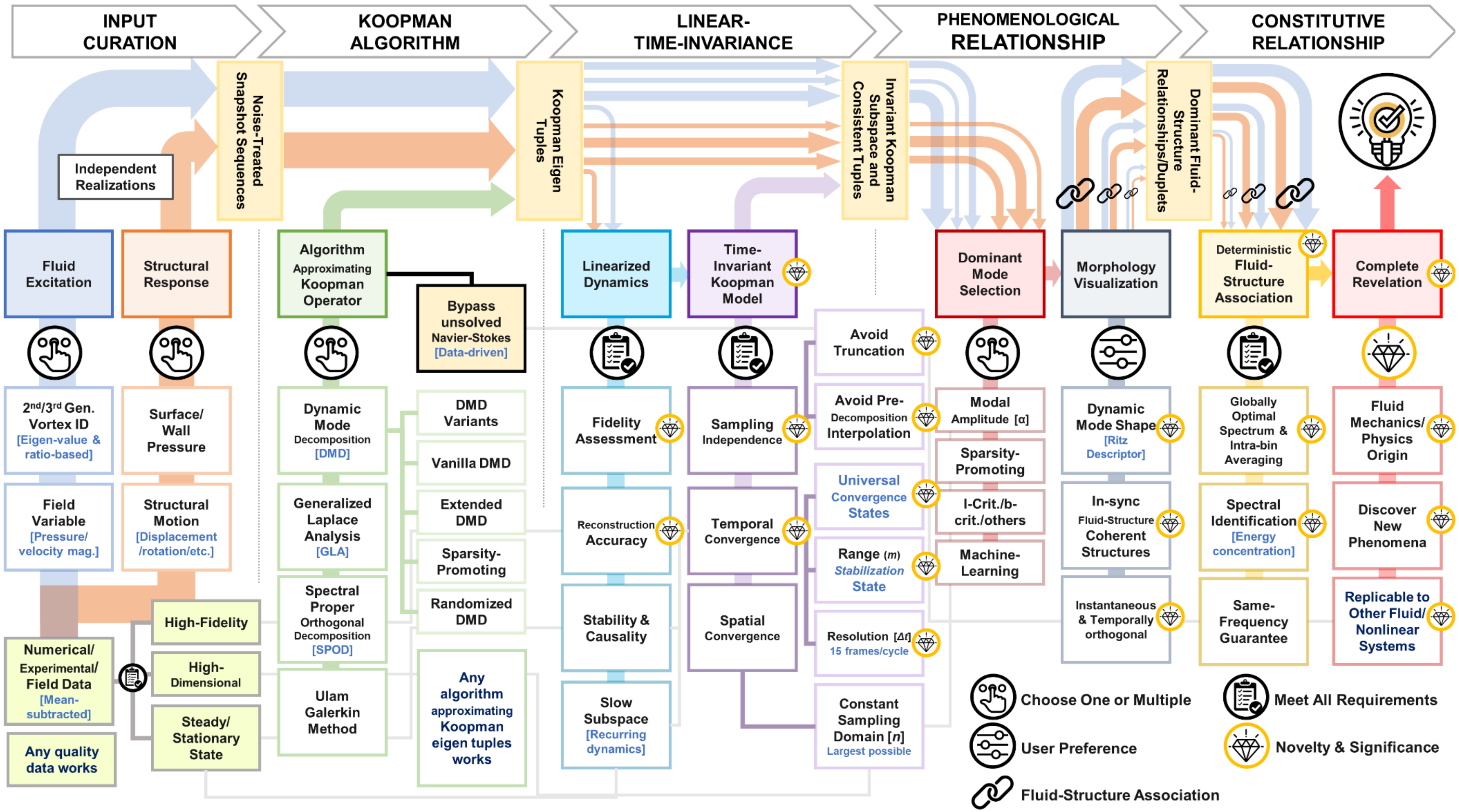

INPUT CURATION
KOOPMAN ALGORITHM
LINEAR-TIME-INVARIANCE
PHENOMENOLOGICAL RELATIONSHIP
CONSTITUTIVE RELATIONSHIP
Independent Realizations
Noise-Treated Snapshot Sequences
Koopman Eigen Tuples
Invariant Koopman Subspace and Consistent Tuples
Dominant Fluid-Structure Relationships/Duplets
Fluid Excitation
Structural Response
Algorithm Approximating Koopman Operator
Bypass unsolved Navier-Stokes [Data-driven]
Linearized Dynamics
Time-Invariant Koopman Model
Dominant Mode Selection
Morphology Visualization
Deterministic Fluid-Structure Association
Complete Revelation
2nd/3rd Gen. Vortex ID [Eigen-value & ratio-based]
Field Variable [Pressure/ velocity mag.]
Surface/ Wall Pressure
Structural Motion [Displacement /rotation/etc.]
Numerical/ Experimental/ Field Data [Mean-subtracted]
Any quality data works
High-Fidelity
High-Dimensional
Steady/ Stationary State
Dynamic Mode Decomposition [DMD]
Generalized Laplace Analysis [GLA]
Spectral Proper Orthogonal Decomposition [SPOD]
Ulam Galerkin Method
DMD Variants
Vanilla DMD
Extended DMD
Sparsity-Promoting
Randomized DMD
Any algorithm approximating Koopman eigen tuples works
Fidelity Assessment
Reconstruction Accuracy
Stability & Causality
Slow Subspace [Recurring dynamics]
Sampling Independence
Temporal Convergence
Spatial Convergence
Avoid Truncation
Avoid Pre-Decomposition Interpolation
Universal Convergence States
Range (m) Stabilization State
Resolution [Δt] 15 frames/cycle
Constant Sampling Domain [n] Largest possible
Modal Amplitude [α]
Sparsity-Promoting
I-Crit./b-crit./others
Machine-Learning
Dynamic Mode Shape [Ritz Descriptor]
In-sync Fluid-Structure Coherent Structures
Instantaneous & Temporally orthogonal
Globally Optimal Spectrum & Intra-bin Averaging
Spectral Identification [Energy concentration]
Same-Frequency Guarantee
Fluid Mechanics/ Physics Origin
Discover New Phenomena
Replicable to Other Fluid/ Nonlinear Systems
Choose One or Multiple
Meet All Requirements
User Preference
Novelty & Significance
Fluid-Structure Association

Figure 1: A schematic illustration of the Koopman Linear-Time-Invariance (Koopman-LTI) Architecture. It consists of the *Input Curation*, *Koopman Algorithm*, *Linearly-Time-Invariance*, *Constitutive Relationship*, and *Phenomenological Relationship* modules. Each module contains several submodules outlining the requirements or options. The Koopman-LTI is data-driven and modular, theoretically accommodating all types of input data and solution algorithms that approximate the Koopman eigen tuples. One may follow this perception and analytical procedure to establish fluid-structure relationships. The in-synch, consistent Koopman modes also bear meaningful implications, enhancing understanding of FSI mechanisms.

| ***Mode* - *Frequency*** | Input Data | | | | | | | | | | | | | | | | | |
|---|---|---|---|---|---|---|---|---|---|---|---|---|---|---|---|---|---|---|
| | **Primary Measurable** | | | | | | | | | **Turbulence** | | | | **Vortex Identification Criteria** | | | | |
| | Wall *(P)* | | | | Flow Field | | | | | Reynolds Stress | | | TKE | 1st Gen. | 2nd Gen. | | 3rd Gen. | |
| | BC | DA | AB | CD | $P$ | $u$ | $v$ | $w$ | $\lvert U\rvert$ | $\langle u'v'\rangle$ | $\langle u'w'\rangle$ | $\langle v'w'\rangle$ | $\langle k\rangle$ | $\lvert\omega\rvert$ | $q$ | $\lambda_2$ | $\Omega$ | $\tilde{\Omega}_R$ |
| $M_1$ - $St_1$=0.1242 | 1 | 1 | 1 | 1 | 1 | 1 | 1 | 26 | 1 | 11 | 20 | 19 | 1 | 1 | 1 | 1 | 1 | 1 |
| $M_2$ - $St_2$=0.1180 | 2 | 2 | 2 | 3 | 2 | 2 | 2 | 11 | 2 | 14 | 19 | 20 | 3 | 2 | 4 | 4 | 3 | 3 |
| $M_3$ - $St_3$=0.2422 | 14 | 13 | 19 | 9 | 5 | 3 | 5 | 42 | 4 | 19 | 40 | 41 | 10 | 4 | 2 | 2 | 2 | 2 |
| $M_4$ - $St_4$=0.1304 | 3 | 3 | 3 | 11 | 3 | 4 | 3 | 31 | 5 | 23 | 21 | 23 | 5 | 8 | 6 | 7 | 5 | 7 |
| $M_5$ - $St_5$=0.0497 | 4 | 4 | 4 | 14 | 4 | 5 | 8 | 1 | 3 | 5 | 8 | 8 | 2 | 3 | 5 | 5 | 7 | 5 |
| $M_6$ - $St_6$=0.0745 | 10 | 17 | 15 | 4 | 8 | 6 | 13 | 4 | 6 | 13 | 12 | 12 | 19 | 5 | 15 | 16 | 14 | 10 |
| $M_7$- $St_7$=0.0683 | 11 | 21 | 17 | 6 | 10 | 7 | 12 | 14 | 8 | 10 | 11 | 11 | 31 | 7 | 8 | 8 | 11 | 9 |
| $M_8$ - $St_8$=0.1428 | 5 | 6 | 5 | 8 | 6 | 8 | 4 | 6 | 7 | 27 | 23 | 22 | 4 | 9 | 17 | 18 | 8 | 11 |
| $M_9$ - $St_9$=0.1739 | 17 | 11 | 21 | 2 | 9 | 9 | 11 | 24 | 11 | 25 | 28 | 27 | 7 | 6 | 3 | 3 | 10 | 8 |
| $M_{10}$ - $St_{10}$=0.1118 | 6 | 5 | 6 | 10 | 7 | 10 | 6 | 25 | 10 | 21 | 18 | 18 | 6 | 16 | 20 | 22 | 13 | 18 |
| $M_{11}$ - $St_{11}$=0.1366 | 7 | 8 | 7 | 18 | 11 | 11 | 7 | 21 | 9 | 24 | 22 | 21 | 23 | 15 | 25 | 28 | 15 | 19 |
| $M_{12}$ - $St_{12}$=0.1056 | 8 | 7 | 8 | 38 | 12 | 13 | 10 | 22 | 13 | 20 | 17 | 17 | 27 | 17 | 31 | 30 | 19 | 24 |
| $M_{13}$ - $St_{13}$=0.1925 | 23 | 27 | 29 | 5 | 13 | 15 | 9 | 28 | 15 | 22 | 31 | 31 | 11 | 11 | 7 | 6 | 6 | 6 |
| $M_{14}$ - $St_{14}$=0.1553 | 9 | 16 | 9 | 24 | 16 | 16 | 20 | 17 | 21 | 26 | 25 | 26 | 21 | 18 | 10 | 11 | 29 | 25 |
| $M_{15}$ - $St_{15}$=0.0559 | 20 | 9 | 10 | 12 | 15 | 19 | 28 | 23 | 16 | 9 | 9 | 9 | 17 | 10 | 26 | 25 | 25 | 21 |
| $M_{16}$ - $St_{16}$=0.3664 | 37 | 33 | 46 | 7 | 25 | 34 | 15 | 64 | 27 | 45 | 59 | 60 | 16 | 39 | 29 | 26 | 23 | 15 |
| $M_{17}$ - $St_{17}$=0.0994 | 24 | 10 | 13 | 29 | 17 | 14 | 21 | 18 | 17 | 18 | 16 | 16 | 12 | 20 | 37 | 37 | 21 | 20 |
| $M_{18}$ - $St_{18}$=0.0373 | 13 | 44 | 20 | 25 | 24 | 20 | 24 | 2 | 14 | 7 | 6 | 6 | 9 | 19 | 34 | 34 | 26 | 29 |
| $M_{19}$ - $St_{19}$=0.0311 | 25 | 30 | 30 | 19 | 29 | 21 | 32 | 3 | 19 | 6 | 5 | 5 | 14 | 13 | 36 | 36 | 40 | 35 |
| $M_{20}$ - $St_{20}$=0.1677 | 26 | 32 | 32 | 21 | 33 | 25 | 22 | 5 | 32 | 28 | 27 | 28 | 32 | 27 | 60 | 58 | 34 | 28 |
| $M_{21}$ - $St_{21}$=0.0807 | 16 | 12 | 11 | 30 | 14 | 12 | 16 | 7 | 12 | 15 | 13 | 13 | 8 | 12 | 16 | 15 | 16 | 14 |
| $M_{22}$ - $St_{22}$=0.0248 | 35 | 20 | 26 | 27 | 30 | 26 | 42 | 8 | 23 | 4 | 4 | 4 | 22 | 21 | 48 | 48 | 41 | 34 |
| $M_{23}$ - $St_{23}$=0.0124 | 33 | 42 | 34 | 43 | 42 | 32 | 41 | 9 | 31 | 2 | 2 | 2 | 26 | 29 | 58 | 63 | 37 | 36 |
| $M_{24}$ - $St_{24}$=0.0435 | 28 | 19 | 27 | 41 | 26 | 17 | 27 | 10 | 20 | 8 | 7 | 7 | 20 | 14 | 23 | 24 | 18 | 17 |
| $M_{25}$ - $St_{25}$=0.0062 | 45 | 39 | 36 | 53 | 43 | 29 | 30 | 12 | 29 | 1 | 1 | 1 | 28 | 37 | 63 | 64 | 38 | 42 |
| $M_{26}$ - $St_{26}$=0.0186 | 34 | 24 | 25 | 36 | 35 | 31 | 34 | 13 | 28 | 3 | 3 | 3 | 25 | 28 | 49 | 52 | 36 | 39 |
| $M_{27}$ - $St_{27}$=0.0621 | 12 | 38 | 16 | 22 | 19 | 22 | 26 | 20 | 22 | 12 | 10 | 10 | 13 | 22 | 30 | 29 | 31 | 31 |
| $M_{28}$ - $St_{28}$=0.4161 | 50 | 40 | 55 | 45 | 52 | 53 | 56 | 67 | 52 | 62 | 66 | 68 | 53 | 43 | 9 | 9 | 54 | 51 |
| $M_{29}$ - $St_{29}$=0.2236 | 27 | 22 | 23 | 15 | 23 | 30 | 14 | 36 | 33 | 37 | 36 | 37 | 34 | 31 | 12 | 10 | 9 | 12 |
| $M_{30}$ – $St_{30}$=0.2484 | 32 | 35 | 37 | 26 | 22 | 27 | 25 | 40 | 25 | 29 | 39 | 39 | 18 | 32 | 22 | 21 | 4 | 4 |

Table 1: Summary of 30 dominant modes and their respective $|\tilde{\alpha}_j|$ ranking in each Koopman-LTI system (Highlighted: 10 most dominant).

## 2. Phenomenological Relationship (Module 5) – Class 1

Before we begin, a limitation of the modal visualization is clarified. Like any other mode shape, an LTI mode only illustrates the bin-wise-averaged relationship of spatiotemporal content, so the mathematical underpinnings are loosely formulated. Therefore, we limit our discussion to the descriptive, phenomenological realm of fluid mechanics. Even so, as the upcoming sections will demonstrate, the information is already immensely rich.

Readers are also reminded that the mode shapes describe only the relative relationship between coherent dynamics. It means the information presented by one mode is quintessentially identical to its opposite-sign counterpart. Therefore, in the subsequent sections, all terms of 'positive' and 'negative' refer to relative correlations instead of any mathematically stringent inter-dependence. According to Lander *et al.* (2016), we also define the *prism base* as the streamwise distance between the downstream wall and *2.5D* and the *near-wake* as the after-wake up to *7D*.

### *2.1 $M_1$ - Shear Layer Dynamics and Bérnard-Kármán Shedding*

#### *2.1.1 Pressure Field P*

We begin with the *Class 1* mechanisms. Figure 2 presents the normalized dynamic Koopman mode $\boldsymbol{M_1}$ ($St_1=0.1242$) of $P$ and the prism walls. The multimedia file depicts the alternating occurrence, development, and shrinkage of separation bubbles adhered to the crosswind walls, as well as the subsequent shedding of coherent wake structures. By frequency-matching, we see the unequivocal in-synch behaviors of the flow field and the corresponding structure reactions.

Specifically, the upstream wall (AB) is stable throughout the shedding cycle. Only near edges A and B, two slivers of extreme pressure alternate in sign. This is the familiar imagery of stagnation and forced separation. On this note, the observation is monotonous for all other modes, with the only difference in the frequency of the sign switch. The observation meets anticipation and explains why the upstream wall contains stationary dynamics compared to its peers, exhibiting the highest growth/decay rate (figure 8 from Part 1).

On the other hand, the other walls' responses are diverse. The sign-alternating bubbles dictate the crosswind responses. Take the top wall (BC) as an example (see figure 2a), when the bubble is barely emerging, an intense pressure band forms near the rear edge C (in blue), which is opposite-sign to the upstream (in red). As the bubble forms and grows, the band becomes increasingly weak. At the climax of bubble intensity, it becomes like-sign with the upstream and even the high-pressure extremity (see figure 3b). Furthermore, the downstream wall (CD) is anti-symmetrically alternating, which is in evident congruence with the bisected architecture of the near-wake. Again, the downstream wall traces back to the behaviors of the separation bubbles, or more primitively, the shear layer dynamics.

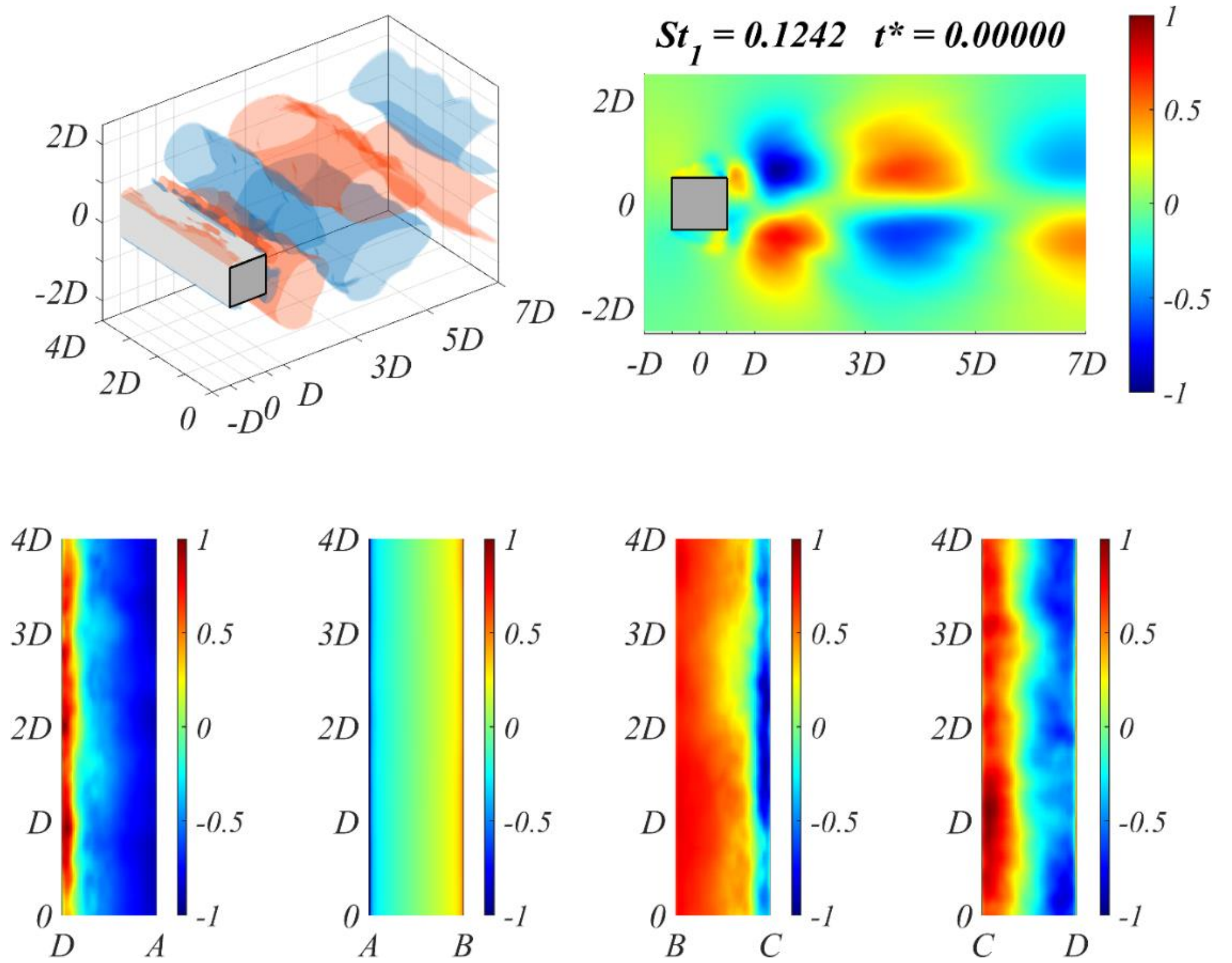

a)

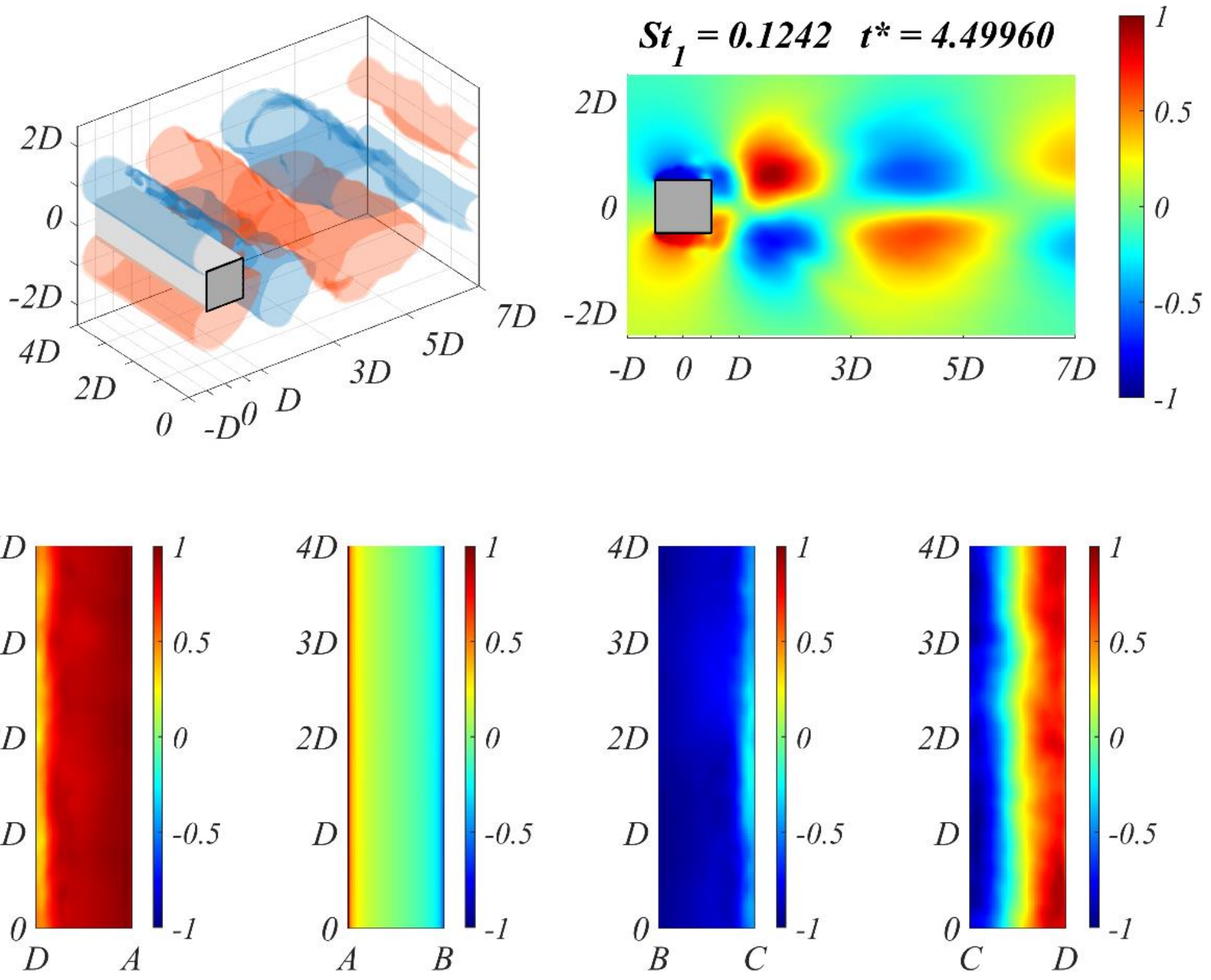

b)

Figure 2: Normalized dynamic Koopman mode *(-1 to 1)* $\boldsymbol{M_1}$ *(*$St_1$*=0.1242)* of *P* and the prism walls at a) *t*=0* and b) *t*=4.49960*: iso-surfaces ±*0.25* of *P* (top left); mid-prism-span slice of *P* (top right); the bottom (DA), upstream (AB), top (BC), and downstream (CD) walls, respectively (bottom from left to right). Multimedia file slowed by a factor of *500*.

### *2.1.2 Velocity Magnitude |U|*

The velocity field is the most common realization and may draw more morphological familiarity. Figure 3 presents the normalized dynamic Koopman mode $\boldsymbol{M_1}$ ($St_1$=*0.1242*) of |*U*| and the prism walls. It delineates two shear layers stemming off the leading edges due to forced separation. Their in-synch motion with *P* confirms that the separation bubbles result from the closure circulation zones, which are directly turbulent without a laminar transition (Kiya & Sasaki, 1983; W. Wu et al., 2020). $\boldsymbol{M_1}$ of |*U*| also illustrates the shear layers' dispersion from

initially intense wall jets into more loosely structured streams as they convect downstream, accompanied by continuous fluid entrainment and vorticity dilution. The shear layers also gain curvature in the process, drawing wake structures increasingly close to the afterbody toward the downstream wall (figure 3b). The ultimate product is the impingement of the leading vortex (Unal & Rockwell, 1988a, 1988b), also known as shear layer reattachment.

The aforenoted wall responses link directly to the shear layers. Take the top wall (BC) as an example again, the pressure band near C propagates counter-streamwise toward B, through which negative pressure turns to positive in a sharp gradient across *1/5D* (see figure 3a). The pressure band results from fluid reversing into the circulating zone, pushing against the forward-traveling mainstream. As the top shear layer disperses and gains curvature, its tendency to close the separation bubble bottlenecks the reverse flow, causing the intensity decay of the band. However, not until the exact moment of closure does the curvature effectuates into actual reattachment (figure 3b). An immediate consequence is the intra-bubble pressure equalization. The low-pressure band also turns high-pressure because the intense fluid of the shear layer stamps onto the wall upon reattachment.

Likewise, the downstream wall (CD) also shows the effect of reattachment. The response is spanwise antisymmetric across the bisecting pressure band. The two halves' pressures are consistently intense and only temporarily relieved as the wake structures cut off their turbulent sheets with the upstream (Sarpkaya, 1979). Only at the exact moment of shedding, the two halves suddenly switch signs and restore the initial intensity (see figure 3b). The consistent pressure results from the curved shear layers approaching the downstream wall. The sign of each side's response also corresponds to that of the respective shear layer.

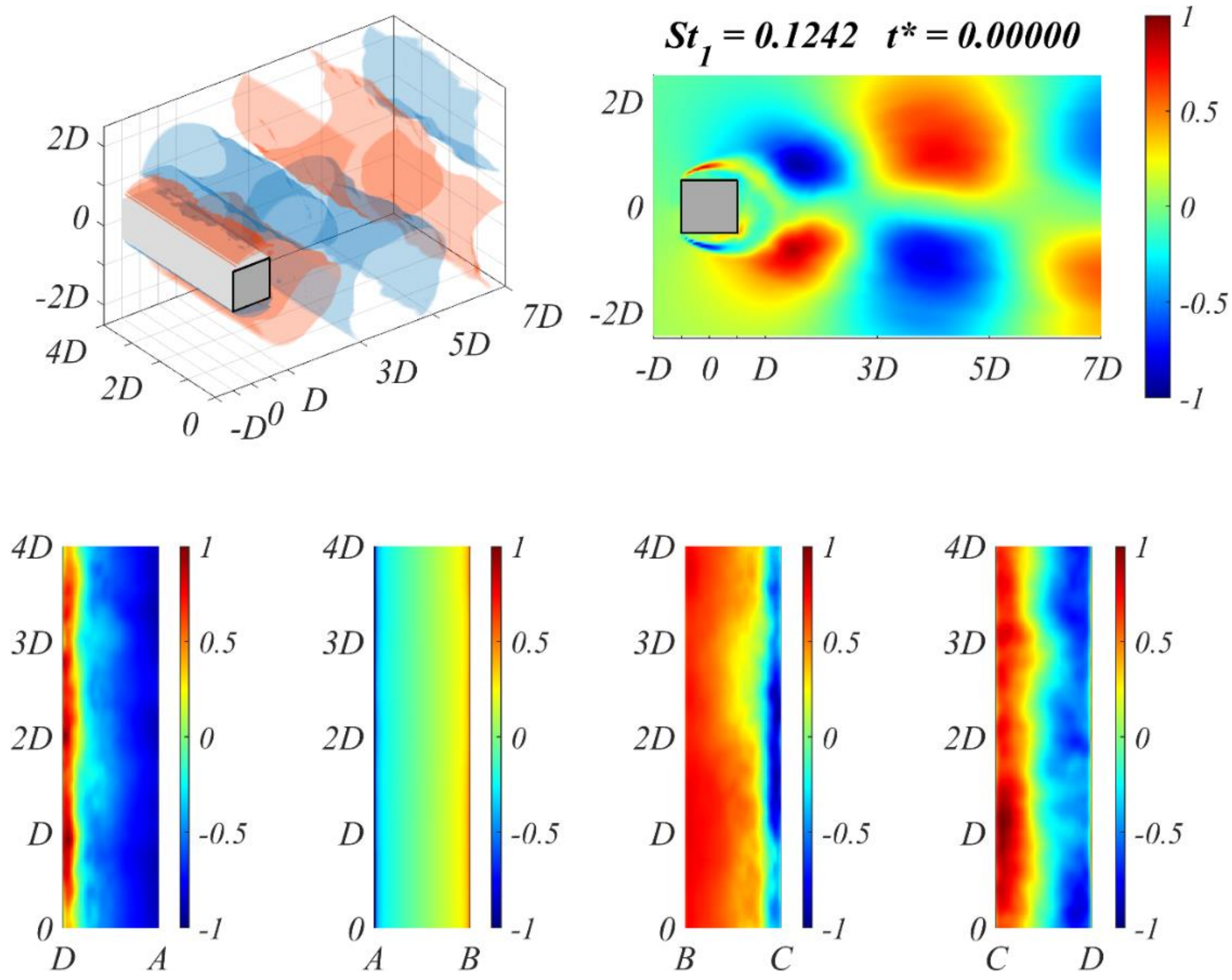

a)

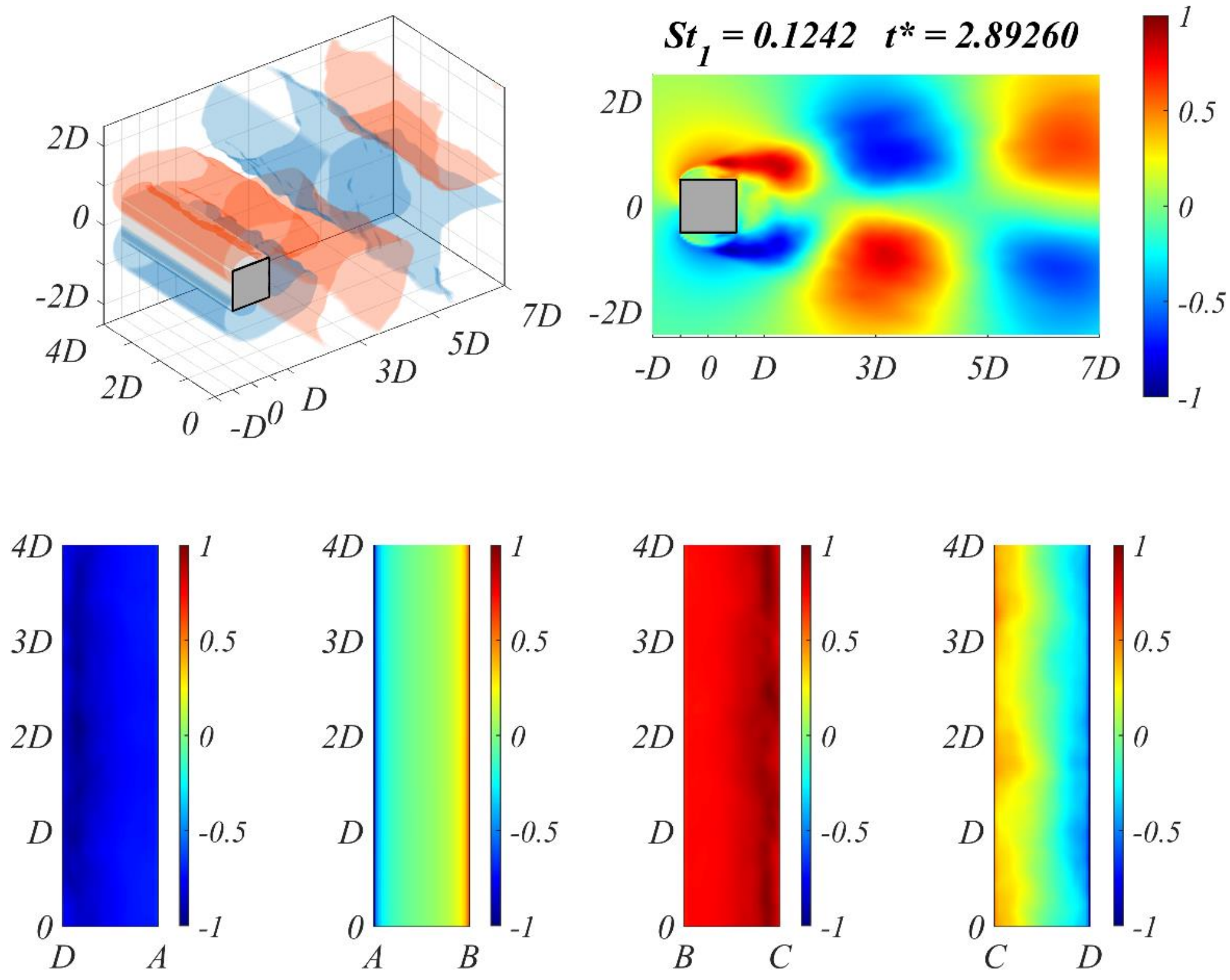

b)

Figure 3: Normalized dynamic Koopman mode *(-1 to 1)* of $\boldsymbol{M_1}$ *($St_1$=0.1242)* of |*U*| and the prism walls at a) *t*=0* and b) *t*=2.89260*: iso-surfaces ±*0.25* of |*U*| (top left); mid-prism-span slice of |*U*| (top right); the bottom (DA), upstream (AB), top (BC), and downstream (CD) walls, respectively (bottom from left to right). Multimedia file slowed by a factor of *500*.

We summarize our observations from $\boldsymbol{M_1}$ into a lucid phenomenological process. A wall jet from forced separation develops at the leading corner and acts as a shear layer. It gains curvature and momentum thickness due to dispersion and entrainment, both of which draw it closer to the afterbody. The curvature increasingly shortens the lateral distance to the rear edge, bottlenecks the reverse flow, and eventually culminates at the exact moment of reattachment. The reattachment, manifested as the closure of separation bubbles, ceases the momentum transfer into the wake by the shear layer. Notwithstanding, the halt of the outlet is not accompanied by that of generation. As such, momentum builds up inside the bubble, dilating it in all directions. The dilation cannot penetrate the wall. It also counters a strong, incessant resistance from the oncoming free-stream, and an even stronger one from the corner-accelerated stream. With no other option, the bubble's membrane-like structure succumbs to the continuous buildup at the feeblest point---the location of reattachment. An immediate consequence is a momentum respite, but at the price of shedding coherent fluid into the prism base. This cyclic process was puzzled together after centuries outstanding research (Bai & Alam, 2018; Cao et al., 2020; Lander et al., 2018; Nakamura & Nakashima, 1986; Portela et al., 2017; Trias et al., 2015; Zhao et al., 2014), and is now effectively isolated and visualized by the LTI.

Dissecting $|U|$ into individual components reveals a consistency between $u$ and $v$, while $w$ appeals to mechanisms unrelated to shear layers. The investigation also supports the overwhelming contribution of $u$ in the content of $|U|$, as their dynamic behaviors display close resemblance (see figure 10 from Part 1). The dynamic Koopman modes of $\boldsymbol{M_1}$ of $u$, $v$ and $w$ are presented in Appendix AI.1-3 to preserve concision.

### *2.1.3 q-Criterion*

A natural step forward is to examine vortex structures in the flow field. This paper presents $q$ and $\tilde{\Omega}_R$ to represent the largely self-similar behaviors of the second-generation and third-generation criteria. We spare the first-generation criterion because Gao & Liu (2018) and Jeong & Hussain (1995) have established critical distinctions between vorticity and a vortex, remarking $|\omega|$ as obsolete. Accordingly, figure 4 presents normalized dynamic Koopman mode $\boldsymbol{M_1}$ ($St_1=0.1242$) of $q$ and prism walls.

Coherent structures are identified alongside the shear layers. They are opposite-sign and outline the momentum thickness, depicting two shearing interfaces between the wall jets and the surrounding flow. This is an elegant picture of the Kelvin-Helmholtz (KH) instability during shear layer transition II (Lander et al., 2016, 2018). We look at the top shear layer as an example. The fluid convects at a low or even opposite velocity inside the recirculation zone, so intense viscous shearing generates the inner interface as the forward- and backward-traveling fluid encounter, causing the rollup of the interfacial KH vortices. By contrast, the outer interface results from the corner-accelerated stream, where fluid convects at high speed. Accordingly, the outer KH structures are opposite-sign to the inner ones, appropriately describing the relative relationship between the two shearing interfaces. Subsequently, the KH structures convect downstream alongside the curved shear layers. Figure 4a also lucidly depicts how the rear corner cuts into the inner interface or a shear layer's momentum thickness. This is the impingement of the leading vortex and unambiguous evidence of reattachment.

On a different note, though $q$ appropriately depicts the KH instability, it fails to identify the wake structures. The failure reflects the issue Liu *et al.* (2019) outlined: the eigenvalue-based criteria highly depend on a user-defined threshold. Finding an appropriate one that suits both the global and local vortex scales is often esoteric. The threshold prescription is even trickier when the spatiotemporal content is transcribed into the Fourier space. Therefore, though $q$, or the second-generation criterion in general, is rich with information, it can be practically intractable, especially after decomposing the vortex field into orthogonal eigen tuples.

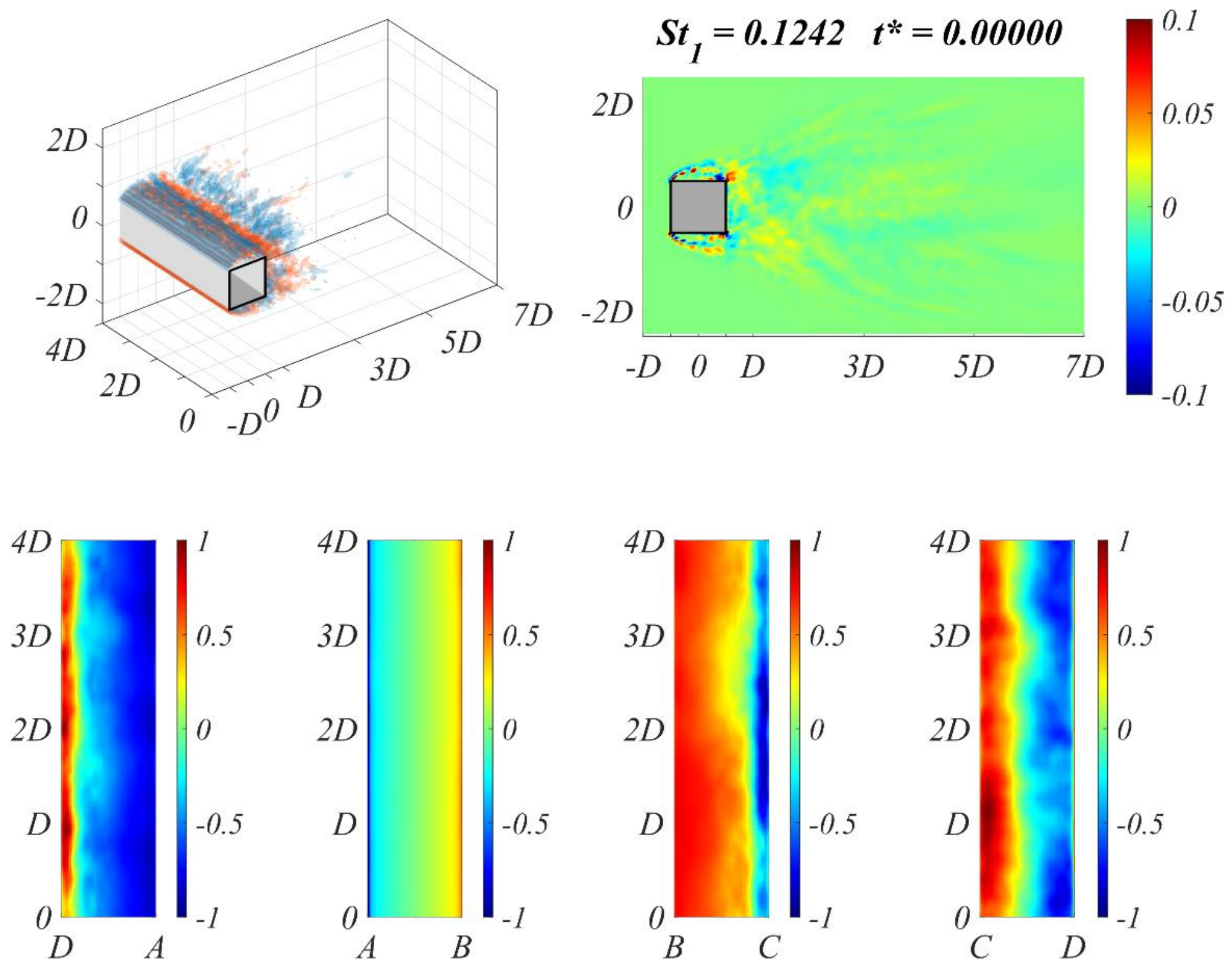

a)

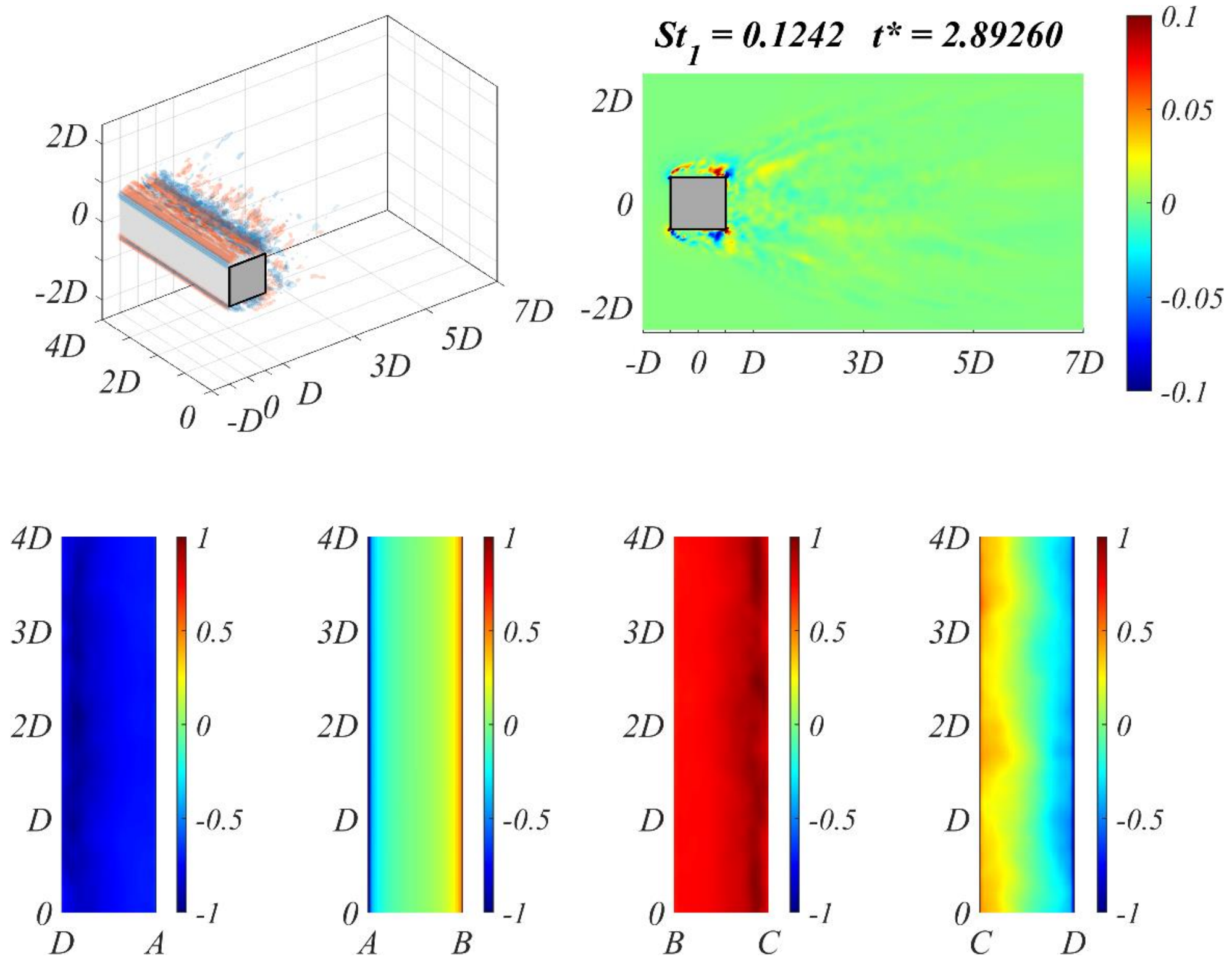

b)

Figure 4: Normalized dynamic Koopman mode *(-1 to 1) of* $\boldsymbol{M_1}$ *(*$St_1$*=0.1242)* of $q$ and the prism walls at a) $t^*=0$ and b) $t^*=2.89260$: iso-surfaces ±*0.25* of $q$ (top left); mid-prism-span slice of $q$ (top right); the bottom (DA), upstream (AB), top (BC), and downstream (CD) walls, respectively (bottom from left to right). Multimedia file slowed by a factor of *500*.

### *2.1.4 $\tilde{\Omega}_R$–Criterion*

Figure 5 presents the normalized dynamic Koopman mode $\boldsymbol{M_1}$ ($St_1=0.1242$) of the third-generation criterion $\tilde{\Omega}_R$ and the prism walls. An immediate observation is its resemblance with the velocity field, reaffirming the coherent structures are associated with vortical activities. However, there is an intriguing catch. A comparison of $|U|$ (figure 3b) and $\tilde{\Omega}_R$ (figure 5b) shows the shedding of the primary structures is escorted by an entourage of small-scale vortices. Interestingly, these smaller vortices are widely scattered in the near wake and do not necessarily conform to the borders of the primary structures. But, as the shedding progresses, the smaller vortices quickly dissipate, and only those within the coherent structures survive (figure 3a and figure 5a). A vortex's rotation acts as the shelter for vorticity.

On a methodical note, the downside of $\tilde{\Omega}_R$ is apparent, too. The price of a universal threshold is the loss of local details. Though still vaguely visible, $\tilde{\Omega}_R$'s delineation of the momentum thickness is less favorable than that of $q$, let alone the more intricate details of the shearing interfaces and KH structures. To this end, the ratio-based criteria are not, at least for the scopes herein, necessarily an *improvement* of the eigenvalue-based criteria. It simply provides an alternative lens to vortex dynamics.

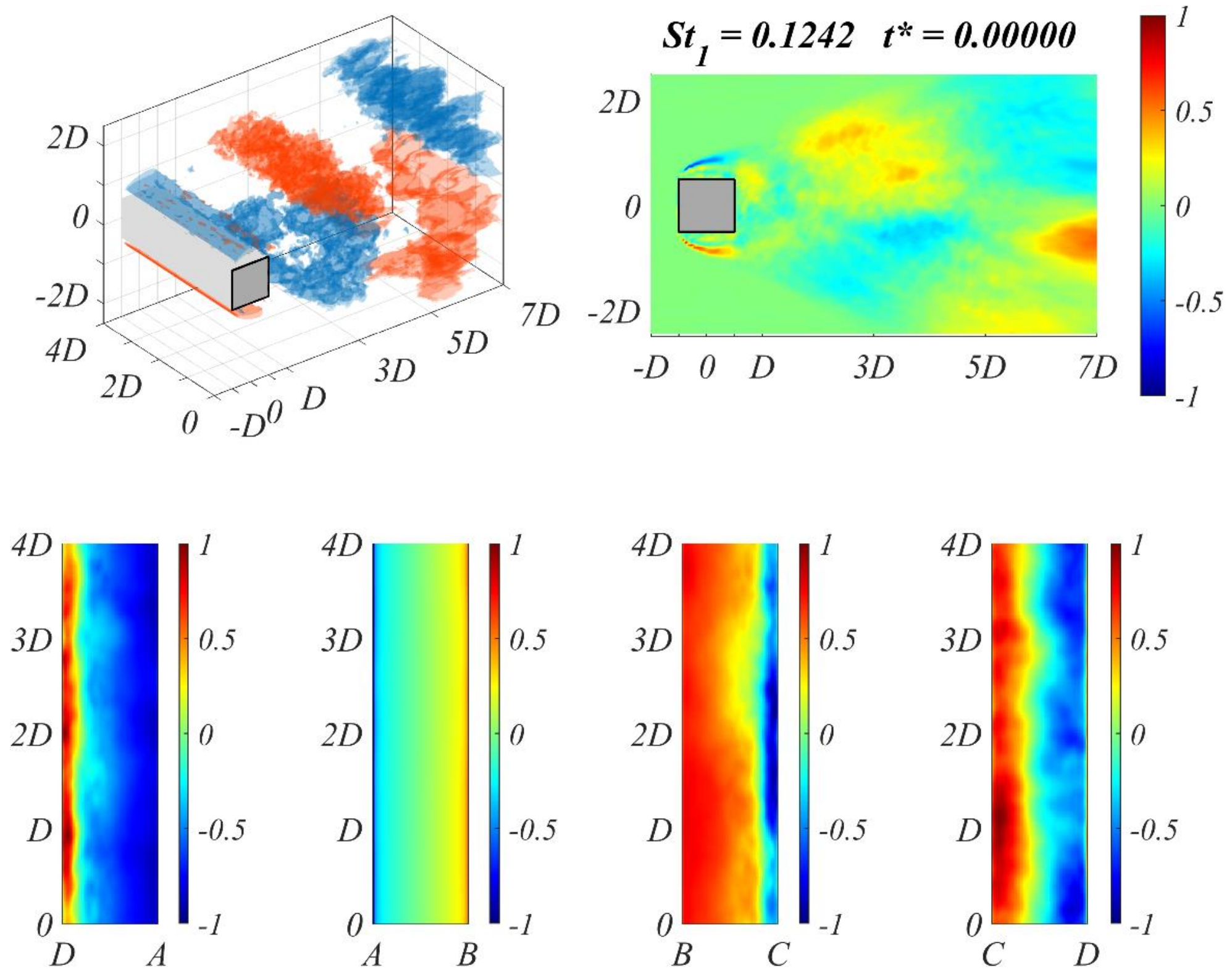

a)

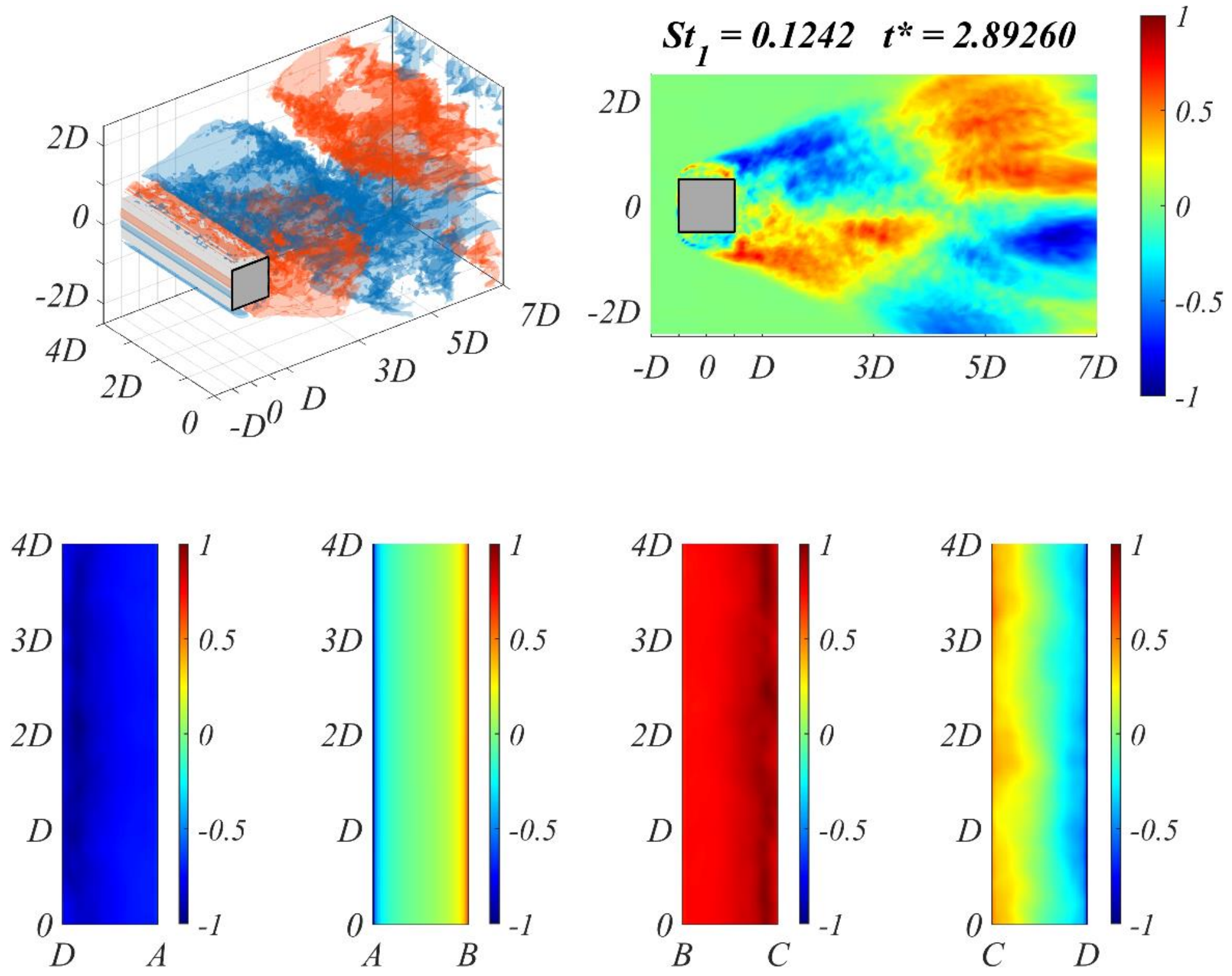

b)

Figure 5: Normalized dynamic Koopman mode *(-1 to 1)* of $M_1$ *($St_1$=0.1242)* of $\tilde{\Omega}_R$ and the prism walls at a) *t*=0* and b) *t*=2.89260*: iso-surfaces ±*0.25* of $\tilde{\Omega}_R$ (top left); mid-prism-span slice of $\tilde{\Omega}_R$ (top right); the bottom (DA), upstream (AB), top (BC), and downstream (CD) walls, respectively (bottom from left to right). Multimedia file slowed by a factor of *500*.

#### *2.1.5 Broadband Content*

Intuition motivates an examination of the broadband content of the primary peak $M_1$. We replicated the same exhaustive procedure for every Koopman mode in the subsequence but presented only the most representative to maintain literary concision. The less pivotal figures are spared or assorted in the appendix. Accordingly, the dynamic Koopman modes of $M_2$ ($St_2$=*0.1180*) and $M_4$ ($St_4$=*0.1304*) of |*U*| are presented in Appendix AI.4-5.

Analysis corroborates that $\boldsymbol{M_2}$ and $\boldsymbol{M_4}$, though less energy-potent, are phenomenologically identical to $\boldsymbol{M_1}$, confirming their broadband origin. The observation also extends to several other modes of adjacent frequencies, namely $\boldsymbol{M_8}$ ($St_8=0.1428$), $\boldsymbol{M_{10}}$ ($St_{10}=0.1118$), $\boldsymbol{M_{11}}$ ($St_{11}=0.1366$), $\boldsymbol{M_{12}}$ ($St_{12}=0.1056$), and $\boldsymbol{M_{14}}$ ($St_{14}=0.1553$). The similitude confirms the conclusion drawn from Part 1, suggesting the broadband content of the primary peak is distributed across several frequency bins about $St=0.1\text{-}0.15$ (see figure 14a from Part 1).

### *2.1.6 Longitudinal Rolls*

$\boldsymbol{M_1}$ and its subsidiaries are related to shear layer dynamics, which results in Bérnard-Kármán shedding's primary structure---the *rolls* (Hussain, 1986). Originating from forced separation, foreshadowed by fluid dispersion and shear layer curvature, instigated by reattachment and shear layers *roll-up* (Li et al., 2022; J. Wu et al., 1996), and supplemented by the intra-shear layer Kelvin-Helmholtz instability (Bloor, 1964; Gerrard, 1966; Khor et al., 2011), vorticity-infused fluid culminate into the span-wise longitudinal rolls, or otherwise known as the Strouhal vortex (J. Wu et al., 1996).

The Koopman-LTI analysis isolated and pinpointed the structure's reattachment-type responses due to the rolls. From raw data, one may now directly tell fluid origin after observing the rear-edge bands. For engineering practice, the reduction of unsteady crosswind lift effectively comes down to the diminution of separation and reattachment. For example, chamfering the leading corners reduces the wall jets' intensity (Kwok et al., 1988), shortening the afterbody length prevents reattachment (Luo et al., 1994; Ongoren & Rockwell, 1988; Zhao et al., 2014), and freestream turbulence weakens separation so the correlations of forces (Lee, 1975; Lyn & Rodi, 1994; McLean & Gartshore, 1992; Vickery, 1966). Fascinatingly, after the preceding analysis, we reckon even chamfering a prism's leading or the trailing corner bear fundamentally different implications in terms of phenomenology.

## 2.2 $M_5$ - Turbulence Production

### 2.2.1 Morphological Substructures

After $\boldsymbol{M_1}$, this section analyzes the secondary peak of *Class 1*, $\boldsymbol{M_5}$. Figure 6 presents the normalized dynamic Koopman mode $\boldsymbol{M_5}$ ($St_5=0.0497$) of $P$ and the prism walls. The morphology notably differs from that of $\boldsymbol{M_1}$, in which two main types of wake substructures are observed (also shown by $\tilde{\Omega}_R$ in Appendix AI.6). The first, denoted as the *tails*, depicts the longitudinal, tail-like coherent structures that appear anti-symmetrically about the wake centerline. The tails cover the entire streamwise distance of the near-wake (figure 6a). The second, denoted as the *blob*, depicts a blob of fluid that adheres to the downstream wall (figure 6b). Interestingly, the tails and blob are separated by an opposite-sign cavity. Structure-wise, the reattachment-type response is still observed on the crosswind walls. Nonetheless, the downstream wall, instead of the symmetric pattern of $\boldsymbol{M_1}$, reflects the overwhelming effect of the blob substructure: the negative pressure at mid-span depends directly on the size and intensity of the wall-adhering fluid.

Apart from $P$, the normalized dynamic Koopman mode $\boldsymbol{M_5}$ of $|U|$, especially figure 7b, unveils a significant revelation. $\boldsymbol{M_5}$ originates from the shear layers, implying $\boldsymbol{M_1}$ and $\boldsymbol{M_5}$ share origin and, therefore, are interrelated. After a comprehensive analysis, we conclude $\boldsymbol{M_5}$ describes turbulence production. The tails are characteristic of the time-averaged production, mainly consisting of shear production with sufficient convection. In most turbulent free-shear flows, the mean velocity gradient and the mean momentum transfer are like-sign, resulting in the positive production that generates the tail structures. This was originally observed by Hussain (1986) from turbulent jets (figure 8a).

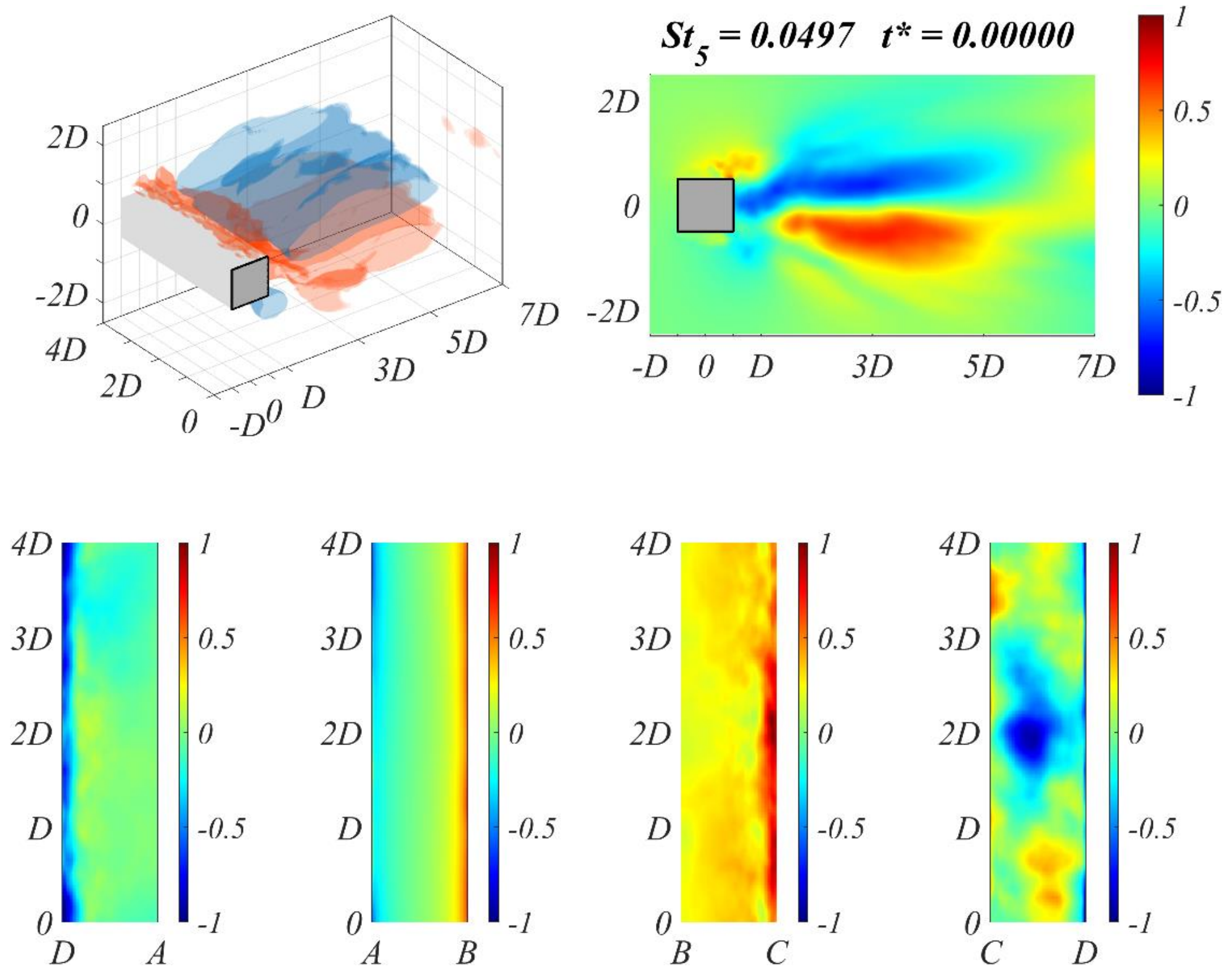

a)

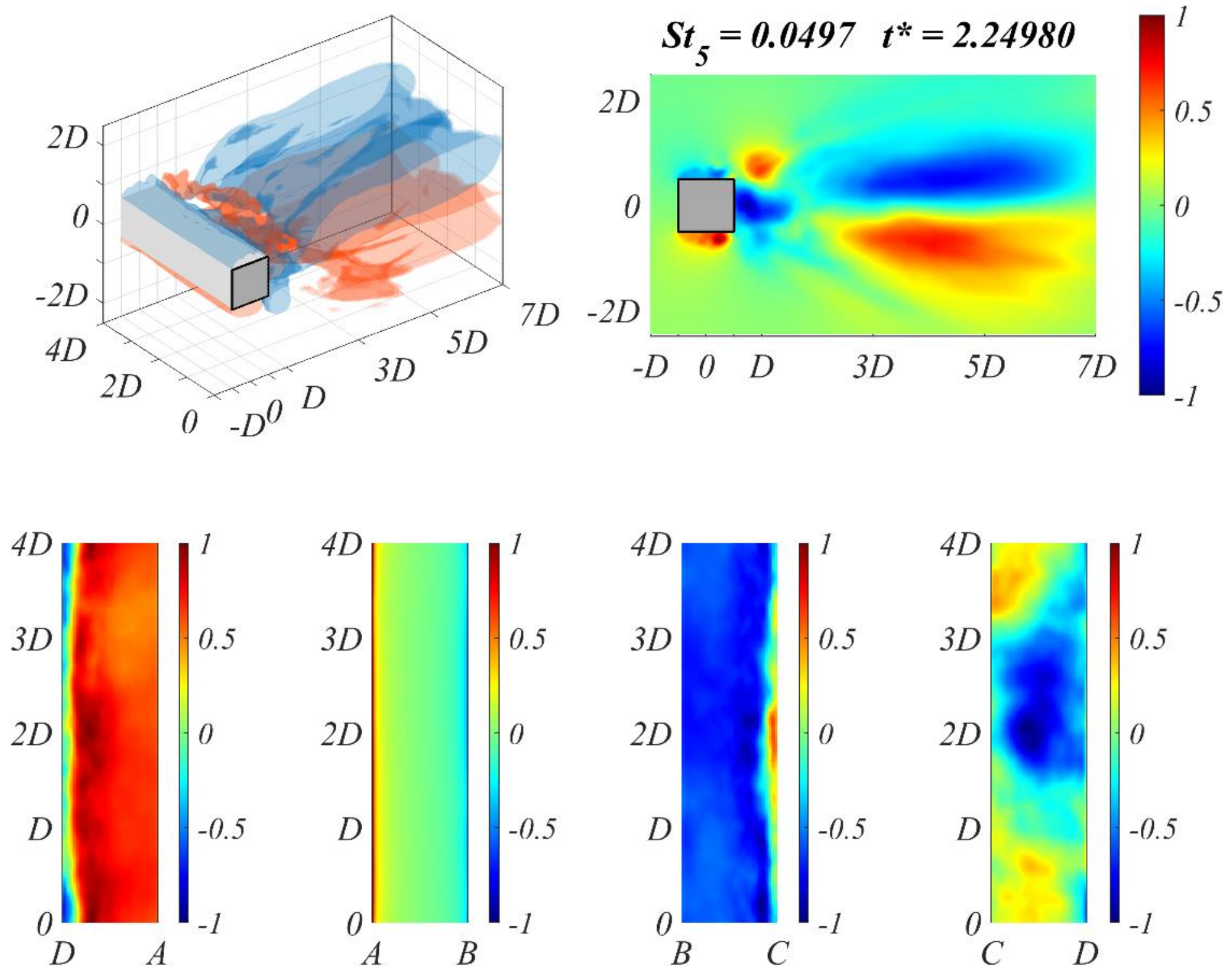

b)

Figure 6: Normalized dynamic Koopman mode *(-1 to 1)* of $\boldsymbol{M_5}$ $(St_5=0.0497)$ of $P$ and the prism walls at a) $t^*=0$ and b) $t^*=2.24980$: iso-surfaces $\pm 0.25$ of $P$ (top left); mid-prism-span slice of $P$ (top right); the bottom (DA), upstream (AB), top (BC), and downstream (CD) walls, respectively (bottom from left to right). Multimedia file slowed by a factor of *500*.

### 2.2.2 Production in the Prism Wake

If one considers the prism wake as two stacked mixing layers or two asymmetric wall jets, then the antisymmetric twin tail morphology strikes as no surprise. However, the shear layers' non-symmetry brings about the issue of negative production, in which the zeros of the mean velocity gradient and the mean momentum transfer do not always coincide. The incongruence produces small regions where the mean velocity gradient and the mean momentum transfer are opposite-sign (figure 8a). The cavity in figure 6b is due to negative production. The dynamic mode shape also animates the cavity's gradual formation and intrusion into the originally one-piece tail structure, cutting it into the detached tail and wall-adhering blob substructures.

The source of production is vortex stretching and fluid entrainment. According to the serial work of Hussain (Hussain, 1981, 1986; Hussain & Hasan, 1985; Hussain & Zaman, 1980), a substructure, known as *ribs*, arises from the stretching of the primary longitudinal rolls. While the shear layers continuously deposit vorticity into the rolls (the process illustrated by $\boldsymbol{M_1}$), the ribs wrap around the rolls in a helical fashion and enrich the longitudinal core's spanwise content. Vortex stretching drives the incessant entrainment of irrotational fluid, and the location of fluid mixing is precisely at the rib-roll interface. This process is like how a ribbed shaft of a grinder draws materials into the machine. However, as the shear layer curves towards the prism base, the negative production isolates the blob from the tail. Consequently, the rib-roll helix is imprinted onto the downstream wall, causing a staggered pattern in which the separatrix of the ribs separates the positive and negative regions. Figure 8b illustrates the rib-roll helix in the prism wake.

In sum, one may trace the shared origin of the *Class 1* mechanisms, namely $\boldsymbol{M_1}$ and $\boldsymbol{M_5}$, to the shear layer dynamics and the associated Bérnard-Kármán Shedding and turbulence production. *Class 1* corresponds to the most natural, energetic field structures and dominate the on-wind wall responses. The dynamical similarity of the on-wind walls also means we can treat them as a spectrally unified fluid-structure interface, despite their geometric disparity. On the other hand, the role of vortex stretching in production can be summarized as the consistent thinning, on statistical average, of fluid elements in the direction perpendicular to the stretching, therefore reducing the radial length scale of the associated vorticity or vortical structures, and ultimately drives the downward cascade into the dissipative scales. Vortex stretching is omnipresent in turbulence.

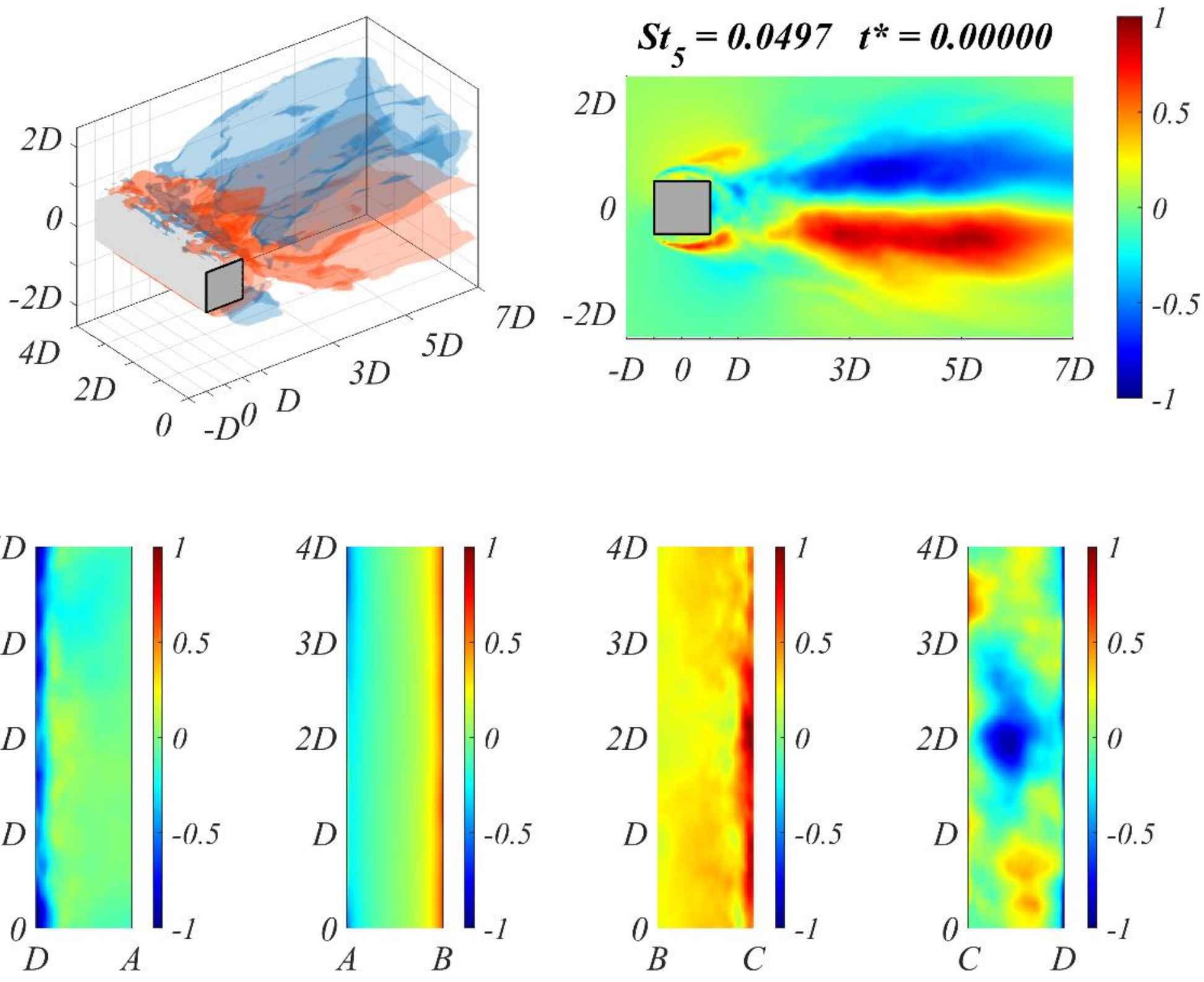

a)

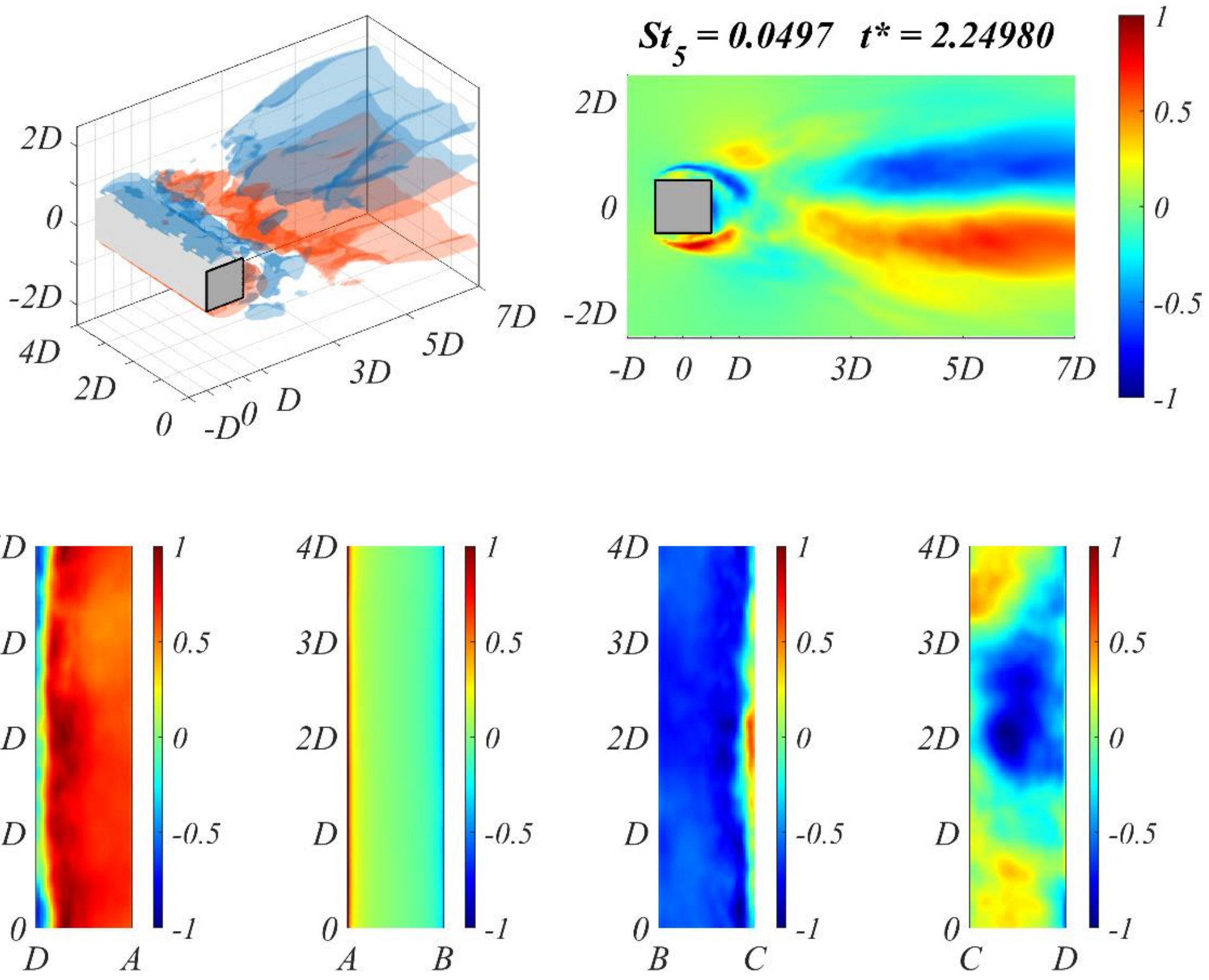

b)

Figure 7: Normalized mode shapes *(-1 to 1)* of $M_5$ *($St_5$=0.0497)* of |*U*| and the prism walls at a) *t*=0* and b) *t*=2.24980*: iso-surfaces ±*0.25* of |*U*| (top left); mid-prism-span slice of |*U*| (top right); the bottom (DA), upstream (AB), top (BC), and downstream (CD) walls, respectively (bottom from left to right). Multimedia file slowed by a factor of *500*.

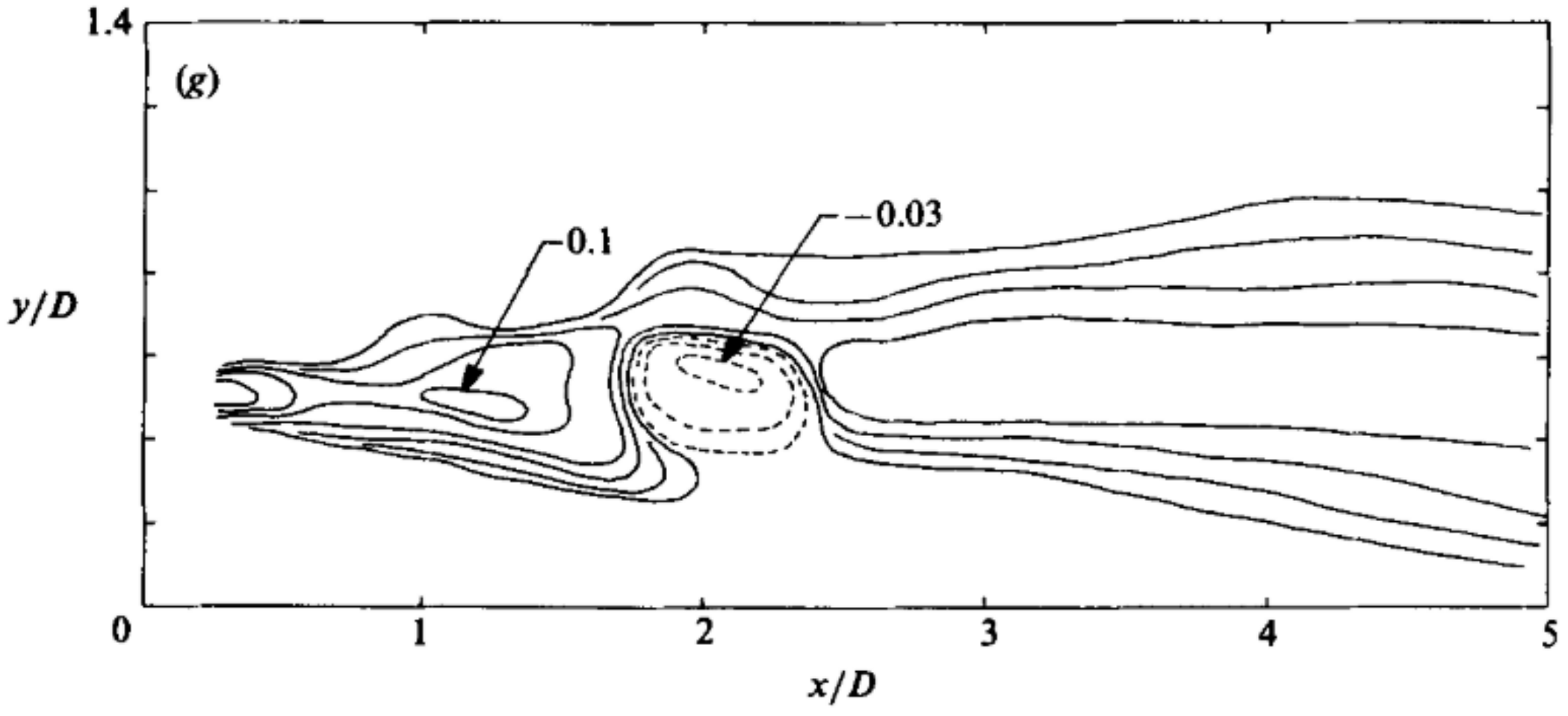

i)

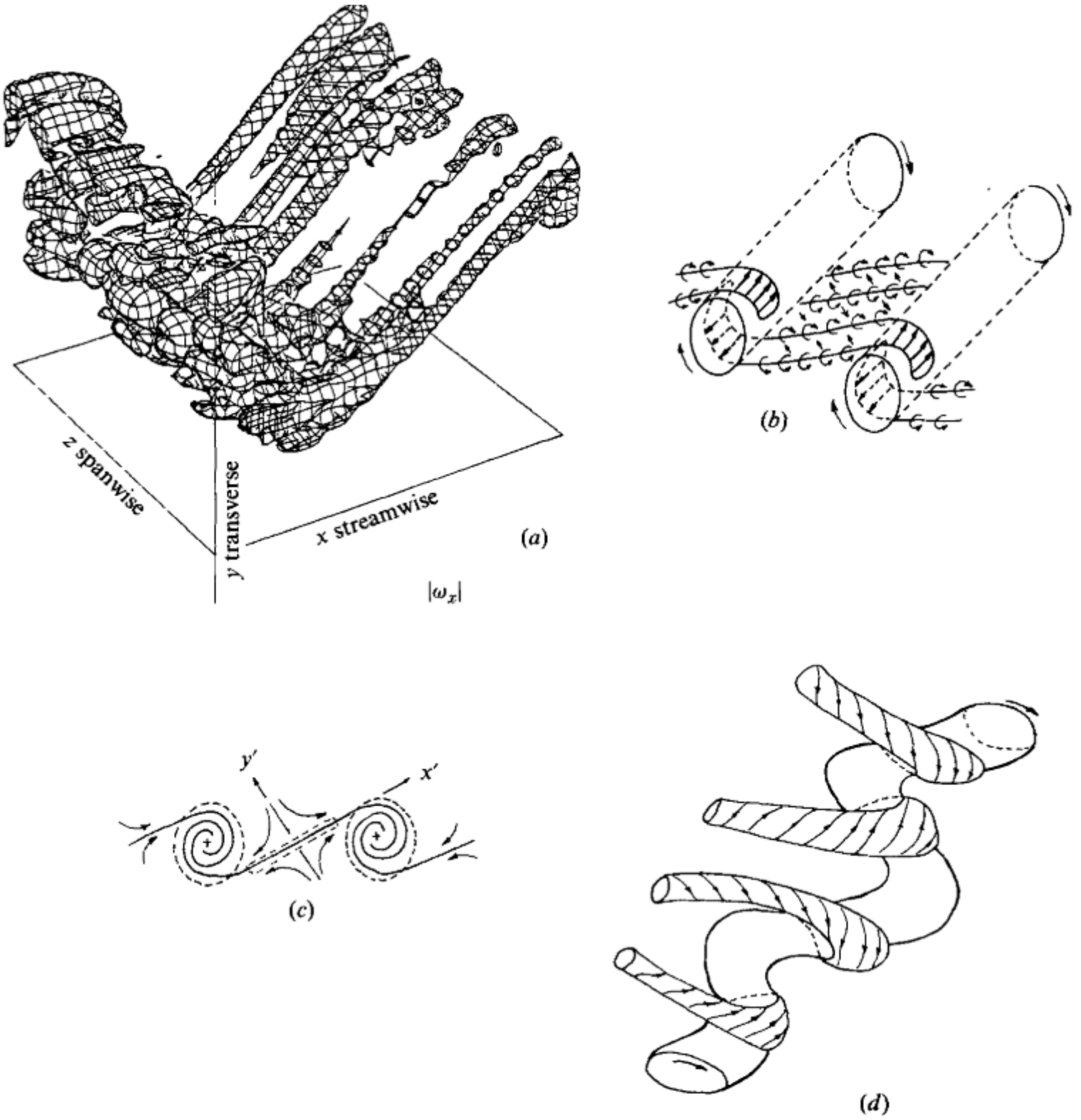

ii)

Figure 8: i) Contour of time-averaged production and negative production (dotted) showing coherent structures in the near field of an axisymmetric jet at the instant of pairing in the jet column mode at *x≈1.75D.* Image taken from figure 3 of Hussain (1986). ii) Dynamics of ribs a) direct numerical simulation, b) schematic, c) flow details around saddle, and d) a more realistic picture of ribs and rolls. Image taken from figure 12 of Hussain (1986). (Included for reviewers' convenience, can be taken out to avoid licensing issues)

## 3. Phenomenological Relationship (Module 5) – Class 2

While the *Class 1* mechanisms overwhelm the on-wind walls, their influence on the downstream wall is far from a monopoly. Part 1 identified four ancillary peaks at $St_3=0.2422$ ($\boldsymbol{M_3}$), $St_7=0.0683$ ($\boldsymbol{M_7}$), $St_9=0.1739$ ($\boldsymbol{M_9}$), and $St_{13}=0.1925$ ($\boldsymbol{M_{13}}$), which overshadow the downstream wall. This section will analyze the phenomenology of the *Class 2* mechanisms.

### 3.1 $M_3$ –Second Harmonic

To begin, figure 9 presents the normalized dynamic Koopman mode $\boldsymbol{M_3}$ ($St_3=0.2422$) of $|U|$ and prism walls. The coherent structures are typical of the widely reported harmonic excitation (Ducoin et al., 2016; Kutz et al., 2016). The frequency $St_3{\sim}2St_1$ confirms $\boldsymbol{M_3}$ is indeed the second harmonic. Though spared from presentation, we also identified higher harmonics like the third harmonic $\boldsymbol{M_{16}}$ ($St_{16}=0.3664$) $St_{16}{\sim}3St_1$. In terms of FSI (see table 1), $\boldsymbol{M_3}$ plays a significant role in the spatiotemporal composition of the flow field (ranks the 4$^{th}$), but its impacts on the on-wind walls are peripheral (ranks the 14$^{th}$, 13$^{th}$, and 19$^{th}$ for BC, DA, and AB, respectively), perhaps only except the downstream wall (ranks the 9$^{th}$). The observation substantiates a rudimentary tenet of fluid-structure interactions---a dominant flow field mechanism does not necessarily incite strong responses from the structure. Even after the global linearization, so the best possible elimination of nonlinearity, FSI is still perplexingly entwined, yet, to an equal degree, bewitchingly fascinating.

As expected, the crosswind walls display the reattachment-type response, conforming to its harmonic lineage. The downstream wall, instead of the mono-band structure of $\boldsymbol{M_1}$, displays a twin-band pattern. For example, in figure 9a, the negative pressure resides in the midspan, and the positive pressures appear near the edges C and D. The opposite pressures are separated by two symmetric bands, doubling that of the fundamental mode for the second harmonic status. This structure response results from an axis-centric wake structure, which detaches the downstream wall starting from the midspan while its two legs linger on the rear edges. This arc oval shape induces a midspan suction with two positive edges (see figure 9a).

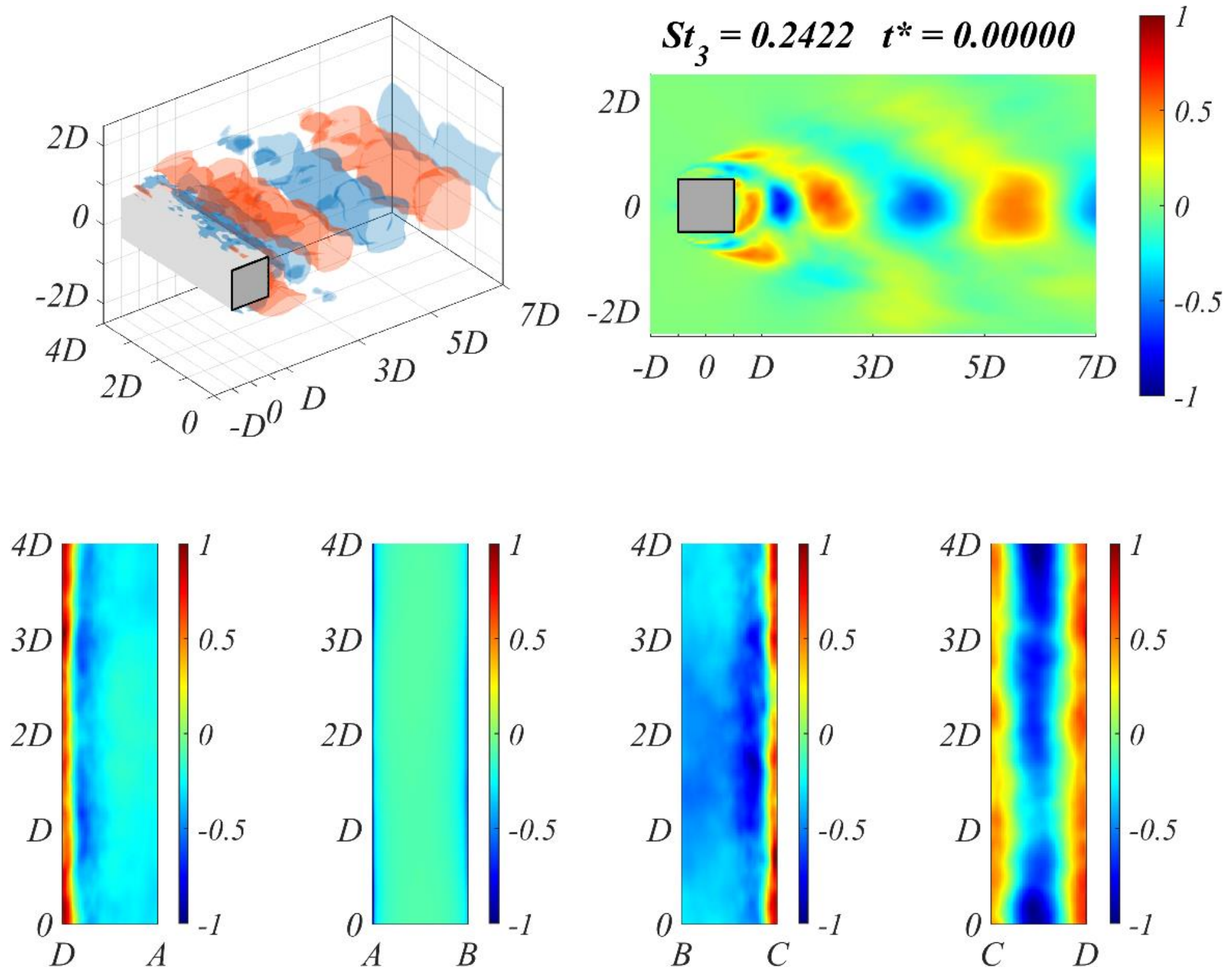

a)

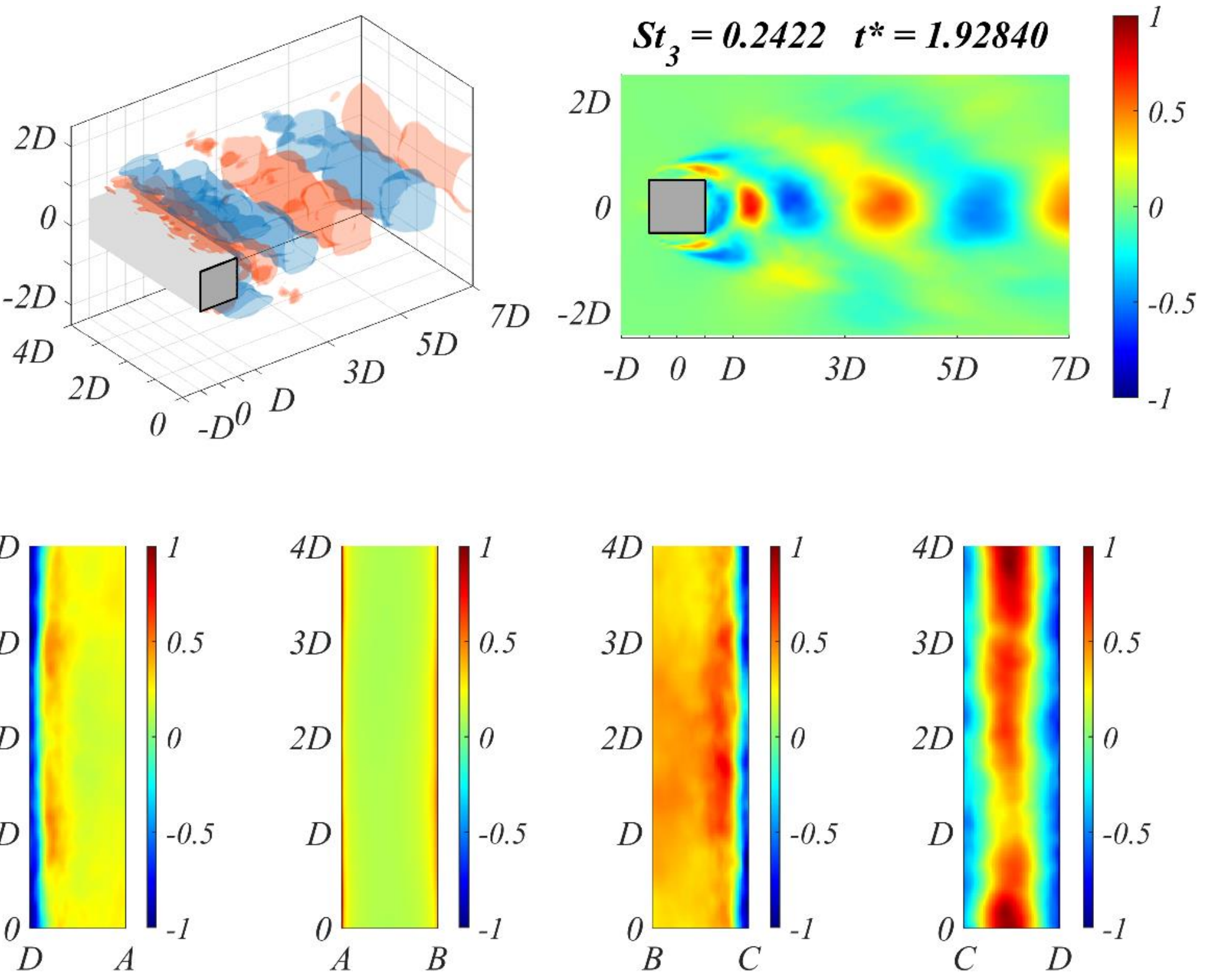

b)

Figure 9: Normalized mode shapes *(-1 to 1)* of **$M_3$** *($St_3$=0.2422)* of |*U*| and the prism walls at a) *t*=0* and b) *t*=1.92840*: iso-surfaces ±*0.25* of |*U*| (top left); mid-prism-span slice of |*U*| (top right); the bottom (DA), upstream (AB), top (BC), and downstream (CD) walls, respectively (bottom from left to right). Multimedia file slowed by a factor of *500*.

### *3.2 $M_7$ –Subharmonic*

Next, figure 10 presents the normalized dynamic Koopman mode **$M_7$** (*$St_7$=0.0683*) of |*U*| and the prism walls. Aside from their frequency, **$M_7$** and **$M_3$** exhibit similarities. **$M_7$**, too, arises from the shear layers. Its coherent structures are axis-centric about the wake centerline (particularly evident in *P* in Appendix AI.7). Like **$M_3$**, **$M_7$** also has a considerable role in the prism wake (ranks the $8^{th}$) but only triggers lukewarm reactions from the prism walls (ranks the $11^{th}$, $21^{st}$, and $17^{th}$ for BC, DA, and AB, respectively), of course, except for the downstream

wall (ranks the 6th). The indifference of the on-wind walls, or the susceptibility of the downstream wall, is in sharp contrast with the *Class 1* mechanisms, indicating $\boldsymbol{M_7}$ appeals to the same class of excitation as $\boldsymbol{M_3}$. For its frequency $St_7 \sim 0.5St_1$, $\boldsymbol{M_7}$ is the subharmonic of the primary structure $\boldsymbol{M_1}$. Our analysis also reveals that $\boldsymbol{M_6}$ ($St_6=0.0745$) is the broadband twin of $\boldsymbol{M_7}$, whose mode shape is merely opposite-sign (Appendix AI.8-9). Ducoin et al. (2016) also made similar observations on the subharmonic peak in the wake of an SD7003 airfoil.

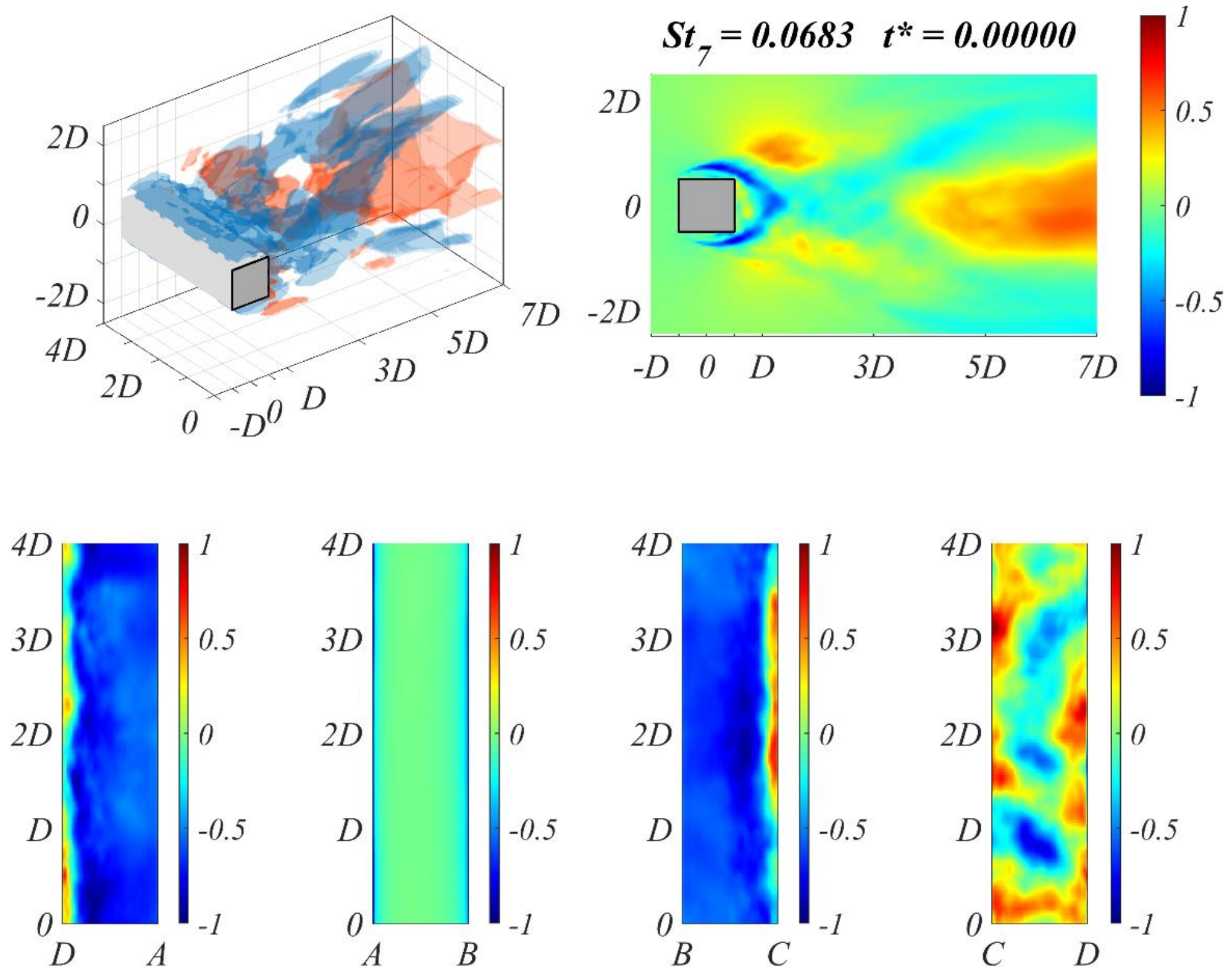

a)

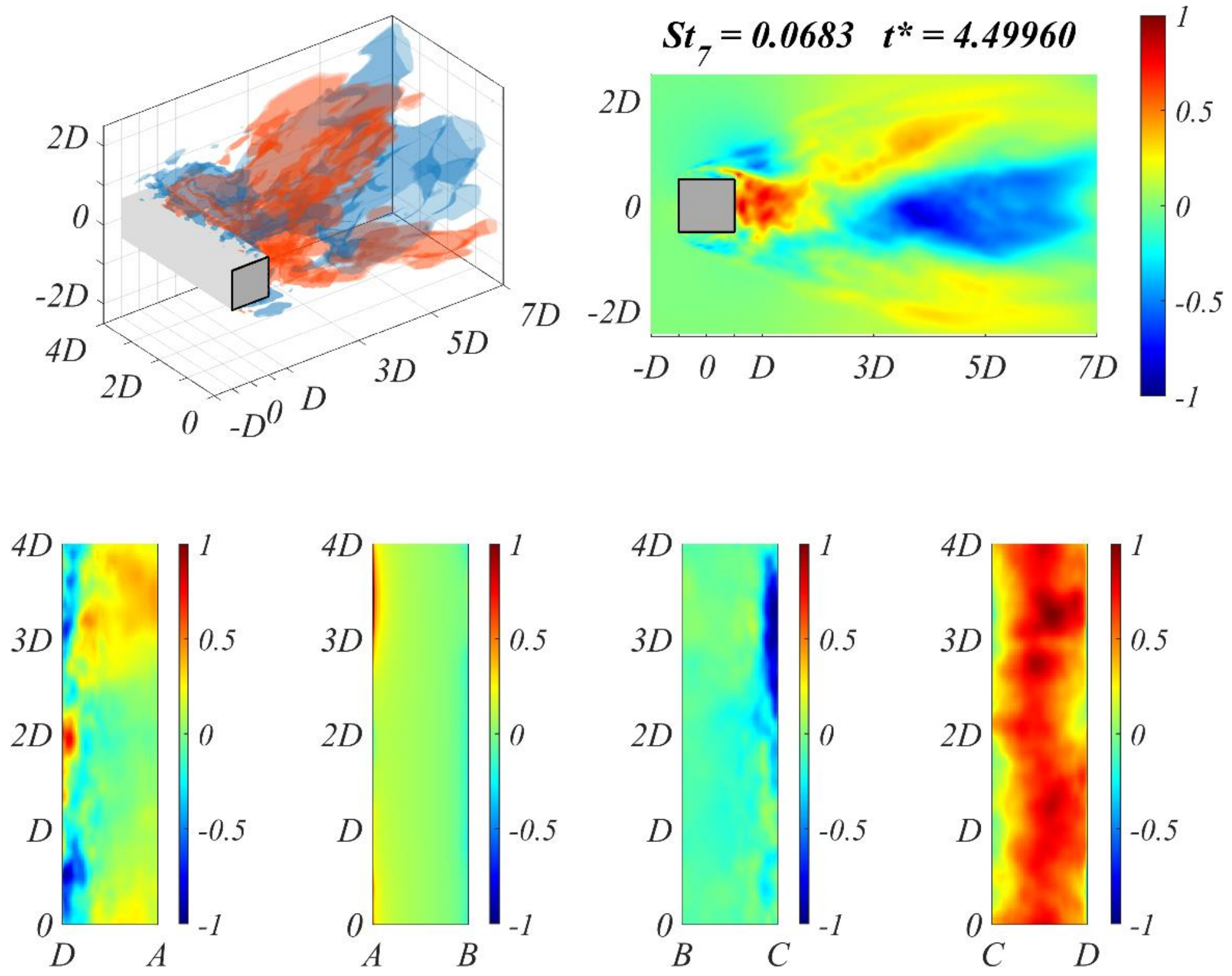

b)

Figure 10: Normalized mode shapes *(-1 to 1)* of $\boldsymbol{M_7}$ *(*$St_7$*=0.0683)* of |*U*| and the prism walls at a) *t*=0* and b) *t*=4.49960*: iso-surfaces ±*0.25* of |*U*| (top left); mid-prism-span slice of |*U*| (top right); the bottom (DA), upstream (AB), top (BC), and downstream (CD) walls, respectively (bottom from left to right). Multimedia file slowed by a factor of *500*.

### *3.3 $M_9$ –Ultra-harmonic*

After pinpointing the second harmonic $\boldsymbol{M_3}$, third harmonic $\boldsymbol{M_{16}}$, and subharmonic $\boldsymbol{M_7}$, a natural sequel is to search for the ultra-harmonic. Without an extenuating effort, we have located $\boldsymbol{M_9}$, which has the frequency $St_9 \sim 1.5 St_1$. Figure 11 presents the normalized dynamic Koopman mode $\boldsymbol{M_9}$ ($St_7=0.1739$) of *P* and the prism walls, illustrating the axis-centric and sequential arrangement of the coherent structures (also |U| in Appendix AI.10). As anticipated, the downstream wall is acutely sensitive to the excitation of the ultra-harmonic. $\boldsymbol{M_9}$, ranking only the 11th in the flow field dominance, incurs the most impactful *Class 2* response, ranking 2nd overall on the downstream wall.

*Class 2* mechanisms root from harmonic excitation and are the culprit of the sophisticated response on the downstream wall. The similarity between the coherent structures of $\boldsymbol{M_3}$, $\boldsymbol{M_7}$, and $\boldsymbol{M_9}$ unilaterally

1) originate from the shear layers with a clear connection to the *Class 1* mechanisms,
2) form in the prism base (<*2.5D*) where negative base pressure is incurred, and
3) remain axis-centric as they convect downstream.

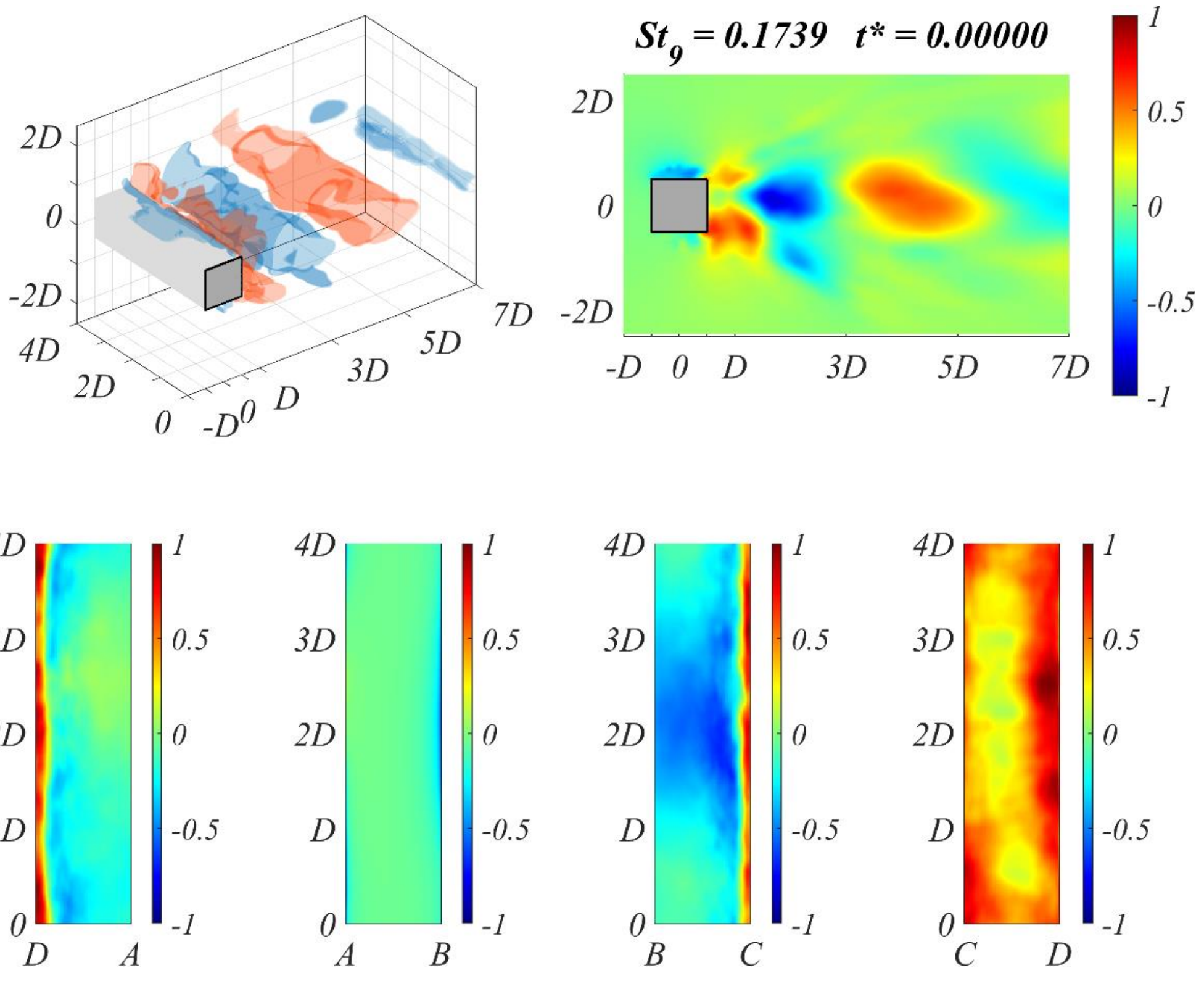

a)

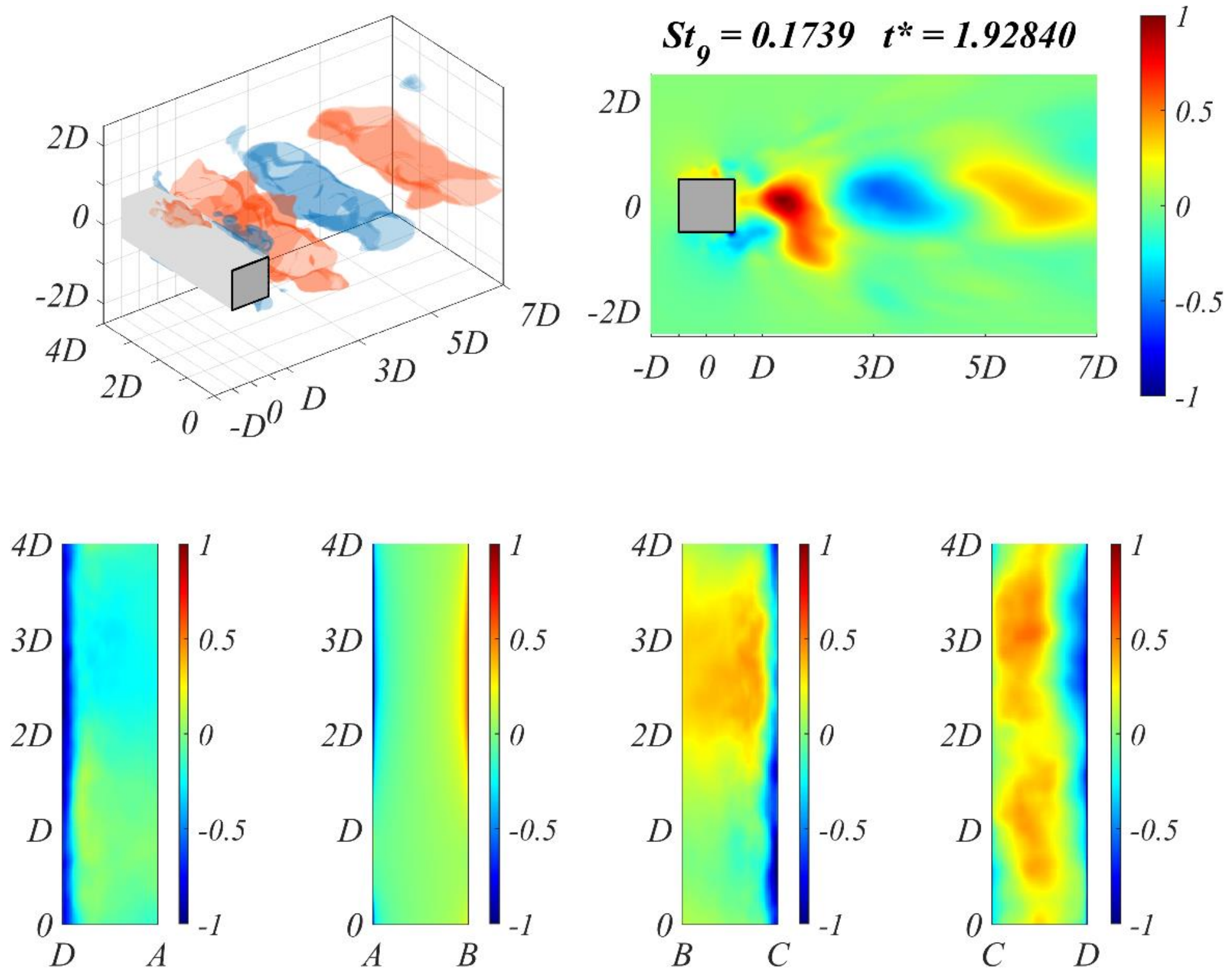

b)

Figure 11: Normalized mode shapes *(-1 to 1)* of $\boldsymbol{M_9}$ *($St_9$=0.1739)* of *P* and the prism walls at a) *t*=0* and b) *t*=1.92840*: iso-surfaces ±*0.25* of *P* (top left); mid-prism-span slice of *P* (top right); the bottom (DA), upstream (AB), top (BC), and downstream (CD) walls, respectively (bottom from left to right). Multimedia file slowed by a factor of *500*.

### *3.4 $M_{13}$ – 2P Mode*

At last, figure 12 presents the normalized dynamic Koopman mode $\boldsymbol{M_{13}}$ (*$St_{13}$=0.1935*) of *P* and the prism walls. $\boldsymbol{M_{13}}$ resembles the other harmonics by dominating the downstream wall. However, it fundamentally differs from others because its coherent structures are not axis-centric. Instead, two parallel, antisymmetric sequences form on either side of the wake axis. The response $\boldsymbol{M_{13}}$ instigates on the downstream wall is also different from its harmonic peers. The mono-band picture appeals to that of the primary structure $\boldsymbol{M_1}$.

To rationalize $\boldsymbol{M_{13}}$, we look into its morphology. Three opposite-sign pairs are found in wake between *0-5D*. In the same region, only two opposite-sign pairs are found for $\boldsymbol{M_1}$. Considering its frequency $St_{13}\sim1.5St_1$, $\boldsymbol{M_{13}}$ is likely a second ultra-harmonic of the fundamental structure. However, unlike the axis-centric sequence of $\boldsymbol{M_9}$, the bi-sequential layout of $\boldsymbol{M_{13}}$ suggests its strong connection with the Kármán Street. More specifically, $\boldsymbol{M_{13}}$'s morphology alludes to the 2P mode originally observed in the wake of vibrating cylinder after surpassing the initial branch (Williamson, 1996; Williamson & Govardhan, 2004).

As described by Williamson & Roshko (1988), the 2P transition is incited when a cylinder's crosswind motion surpasses a critical value, generating a phase difference between two sub-vortices in a single shedding cycle. The phase difference prevents like-sign vortex amalgamation, hence 2-Pairs (2P) of vortices instead of 2-Single (2S) ones form in the Kármán Street. On this note, we must highlight the differences between the test subject in Williamson & Roshko (1988) and herein, which are aeroelastic versus stiff, and cylinder versus prism.

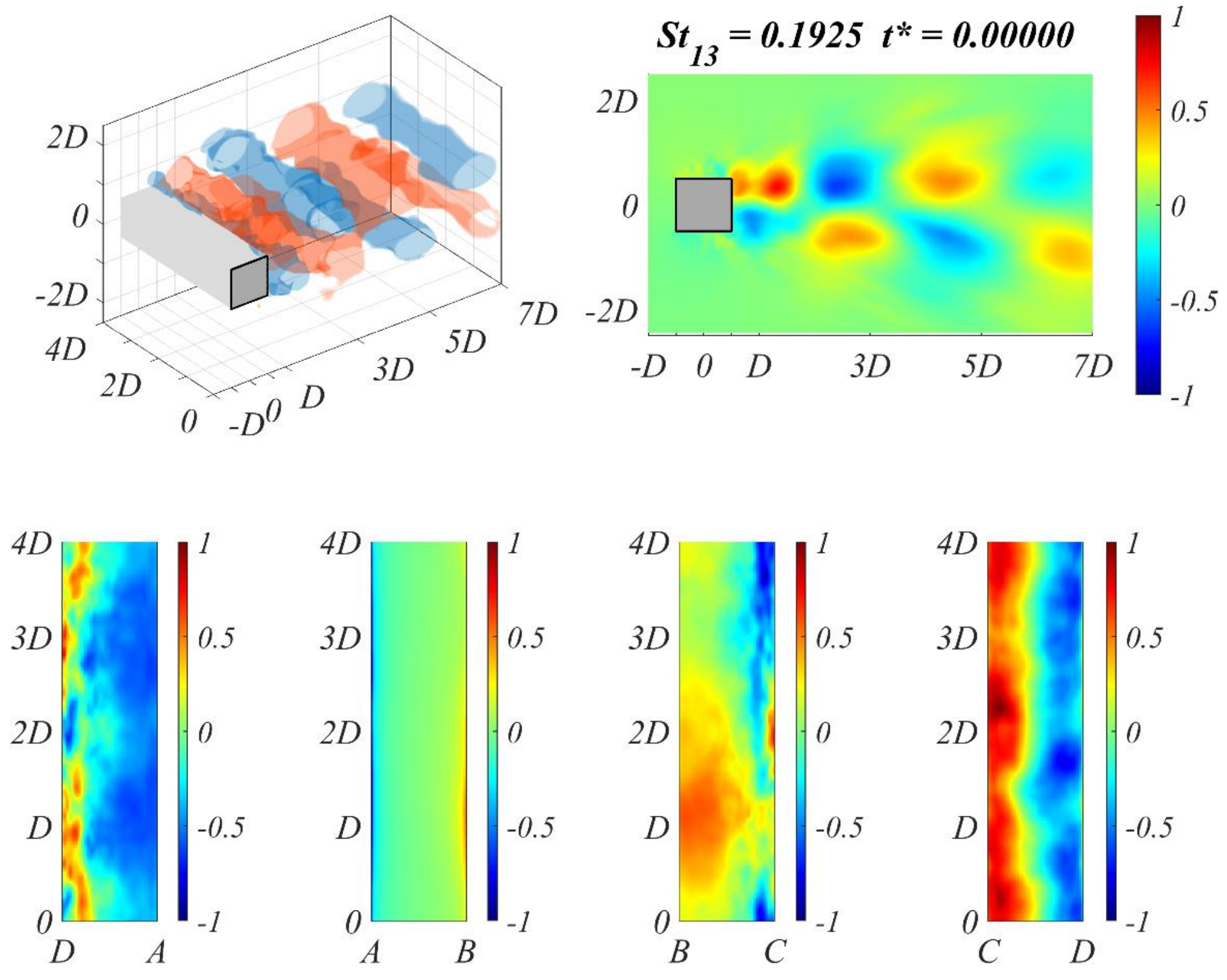

a)

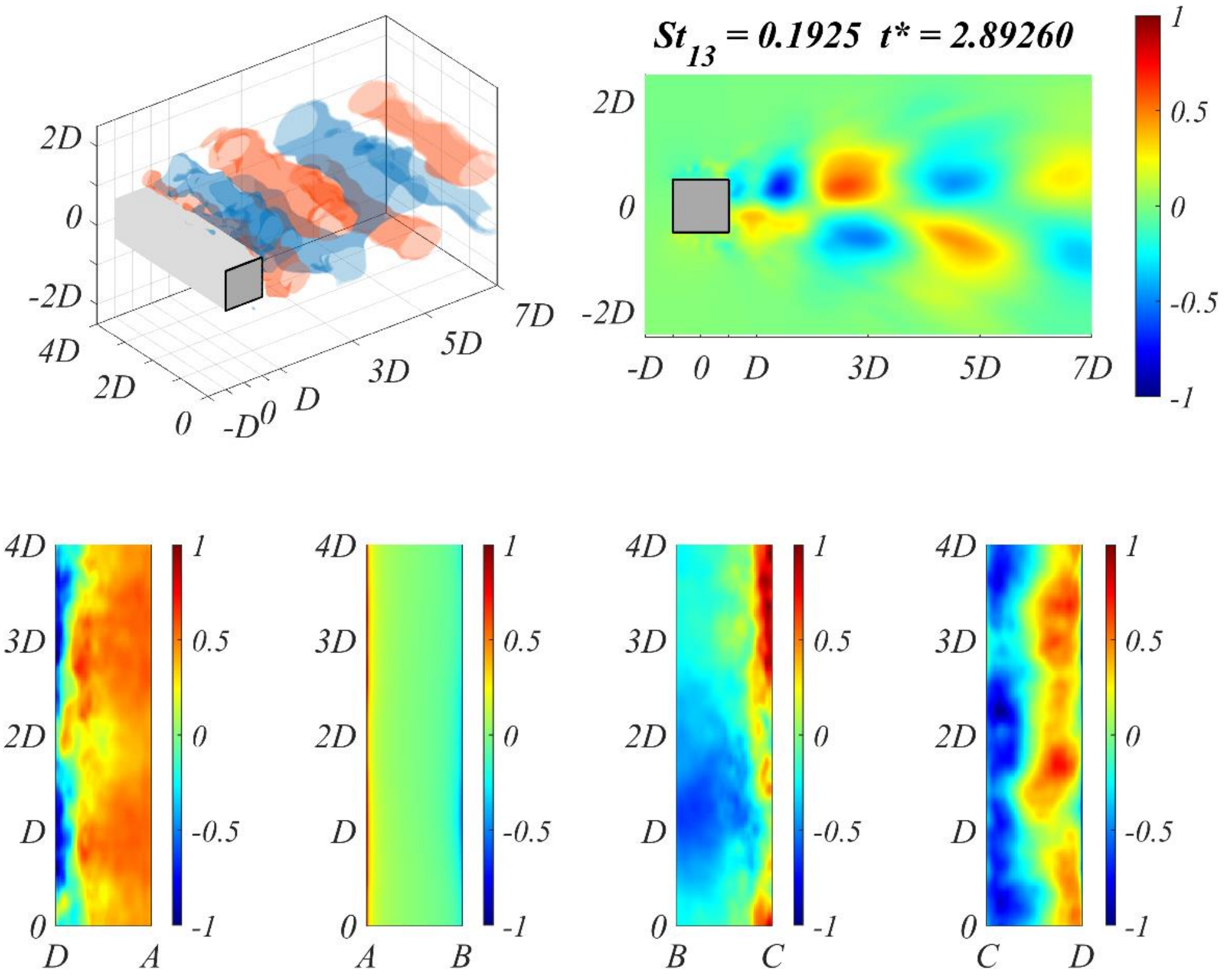

b)

Figure 13*:* Normalized mode shapes *(-1 to 1)* of $\boldsymbol{M_{13}}$ ($St_{13}=0.1925$) of *P* and the prism walls at a) $t^*=0$ and b) $t^*=2.89260$: iso-surfaces ±*0.25* of *P* (top left); mid-prism-span slice of *P* (top right); the bottom (DA), upstream (AB), top (BC), and downstream (CD) walls, respectively (bottom from left to right). Multimedia file slowed by a factor of *500*.

Nevertheless, is it possible that the 2P mode is a natural ultra-harmonic structure of the bluff body wake? One may rationalize how an oscillating cylinder generates the phase difference. A cylinder's lack of a sharp-edged afterbody means it must rely on crosswind motion to prematurely break the turbulent sheet before vortex amalgamation. If the motion is not substantial enough, the wake remains in the preferred 2S state. However, in the reference frame of the cylinder, the only difference between the stiff and aeroelastic scenarios is the curvature of the shear layers, such that the greater the oscillation, the more curved the shear layers, the earlier the reattachment.

What if other mechanisms can enhance the curvature for premature reattachment to the same effect? The shortening of the formation length with *Re* is widely known for the prism wake (Gerrard, 1966; Williamson & Govardhan, 2004), which is indeed due to increasingly curved shear layers. Compared to a curvilinear cylinder, a prism's sharp-edged afterbody incisively cuts turbulent sheets when reattachment takes place. This means structural oscillation, as a way to encourage reattachment, can be spared. Therefore, a phenomenological possibility exists for the premature shedding of the phase-shifted sub-vortices. If so, the 2P shedding in the static prism wake is a spectrally embedded modal behavior. Finally, the layouts of the other harmonics display a striking resemblance with that of the 2S mode (Morse & Williamson, 2009; Williamson & Govardhan, 2004)---axis-centric, alternating, and vividly mono-sequential. Further investigation on this issue will be an exciting exploration.

# 4. New Phenomenon: Vortex Breathing

At this point, we successfully underpinned the fluid origins of the six dominant excitation-response mechanisms. The dynamic Koopman mode shapes described the pedagogical prism wake's phenomenology with accuracy and insights. This paper also demonstrated the methodical procedure to arrive at the conclusions, which is replicable to other flows. One may also extend the conclusions to practical benefit: users can now target a specific fluid phenomenon to eliminate an undesired structural response. For example, one can use a splitter plate to prevent the axis-centric harmonic excitations, thus eliminating pressure extremities on the downstream wall (Song et al., 2017; Unal & Rockwell, 1988b). On this note, we emphasize again that this work stands on the brilliance and effort of centuries of fluid mechanics explorations.

## *4.1 Vortex breathing*

Besides the complete revelation, we now demonstrate a newly discovered phenomenon. We detected an intriguing feature of the detached wake structures by the dynamic mode shape. Take the multimedia file of figure 2 as an example, the coherent structures decay in intensity immediately after breaking their turbulent sheets with the separation bubbles. The decay is dissipation-wise natural and underlined by the contraction of the iso-surfaces and the fading of the midplane contour between *D-3D*. But surprisingly, the structures expand in the follow-up between *3D-5D*. This contraction-expansion motion repeats itself as if the vortices are inhaling and exhaling, hence the *vortex breathing*.

The vortex breathing is fascinatingly perplexing because it disobeys intuitions. If the initial intensity decay is related to the inter-molecular viscous dissipation, then the subsequent growth is unaccounted for. Nonetheless, when adding all the modes together, the energy decays as anticipated. We attribute the breathing phenomenon to the energy exchange in and out of the discrete Koopman modes. Given the difference in periodicity, the inhale of one mode corresponds to the exhale of some others. This exchange of modal energy is an accurate reflection of the wake's dynamical nature. A vortex's downstream convection incessantly injects vorticity into the irrotational fluid in its path, reeling them into circulation. It, too, constantly deposits viscously dissipated fluid in its trail. Therefore, there is a constant energy exchange in the circulation-entrainment-deposition process, which is captured by the energy in and out of a specific eigenfrequency. This also explains why oscillatory sinusoids are excellent descriptors of the wake dynamics (see figure 8 in Part 1).

On a methodical note, vortex breathing demonstrates the importance of the dynamic mode shape. From **figures** 3a and 3b alone, the coherent structures between *3D* to *5D* appear, and quite naturally, less intense, less compact, and more dispersed compared to their successors between *D* to *3D*. The breathing motion would have been effortlessly swept under the carpet by a static DMD mode. Conversely, a logical explanation would have been extremely difficult if the static snapshot is taken at the moment of exhaling, in which the downstream vortex appears more energetic.

# 5. Conclusions

This serial effort focused on the analytical end of the Koopman analysis. We proposed a linear-time-invariance (LTI) notion, or the Koopman Linearly-Time-Invariant (Koopman-LTI) modular architecture, to relate the fluid excitation and structure response. The LTI models reduced the pedagogical prism wake undergoing the shear layer transition II to only six dominant excitation-response mechanisms in Part 1. This Part 2 dynamically visualized their mode shape and revealed their phenomenology, offering a complete revelation of the prism wake.

Specifically, two dynamic Koopman modes at $St_1=0.1242$ and $St_5=0.0497$ describe shear layer dynamics and the associated Bérnard-Kármán Shedding and turbulence production, which overwhelm the upstream and crosswind walls by instigating a reattachment-type of response. The dynamical similarity of the on-wind walls also means we can treat them as a spectrally unified fluid-structure interface, despite their geometric disparity. Another four harmonic counterparts, namely the subharmonic at $St_7=0.0683$, the second harmonic at $St_3=0.2422$, and two distinct ultra-harmonics at $St_7=0.1739$ and $St_{13}=0.1935$, dominate the downstream wall and only marginally affect the others. The 2P wake mode is also observed as an embedded harmonic of the bluff-body wake. This paper also proposed the dynamic Koopman mode, through which we discovered the vortex breathing phenomenon describing the constant energy exchange in wake's circulation-entrainment-deposition processes.

## Acknowledgements

We give a special thanks to the IT Office of the Department of Civil and Environmental Engineering at the Hong Kong University of Science and Technology. Its support for installing, testing, and maintaining our high-performance servers is indispensable for the current project.

# Appendix I

### *AI.1 $M_1$ - u*

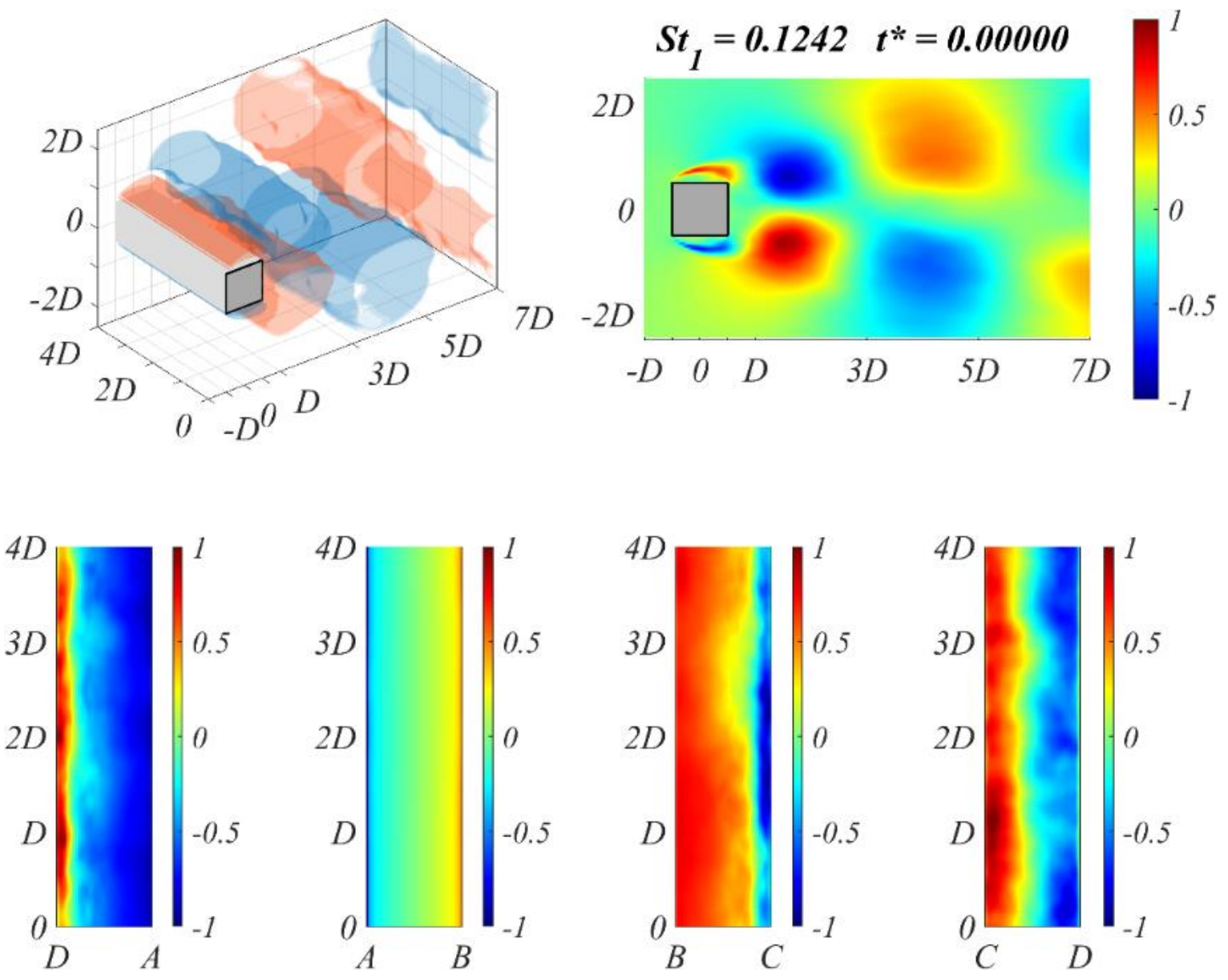

AI. 1: Normalized mode shapes *(-1 to 1)* of $\boldsymbol{M_1}$ *($St_1$=0.1242)* of *u* and the prism walls: iso-surfaces ±*0.25* of *u* (top left); mid-prism-span slice of *u* (top right); the bottom (DA), upstream (AB), top (BC), and downstream (CD) walls, respectively (bottom from left to right). Multimedia file slowed by a factor of *500*.

## *AI.2* $M_1$ *- v*

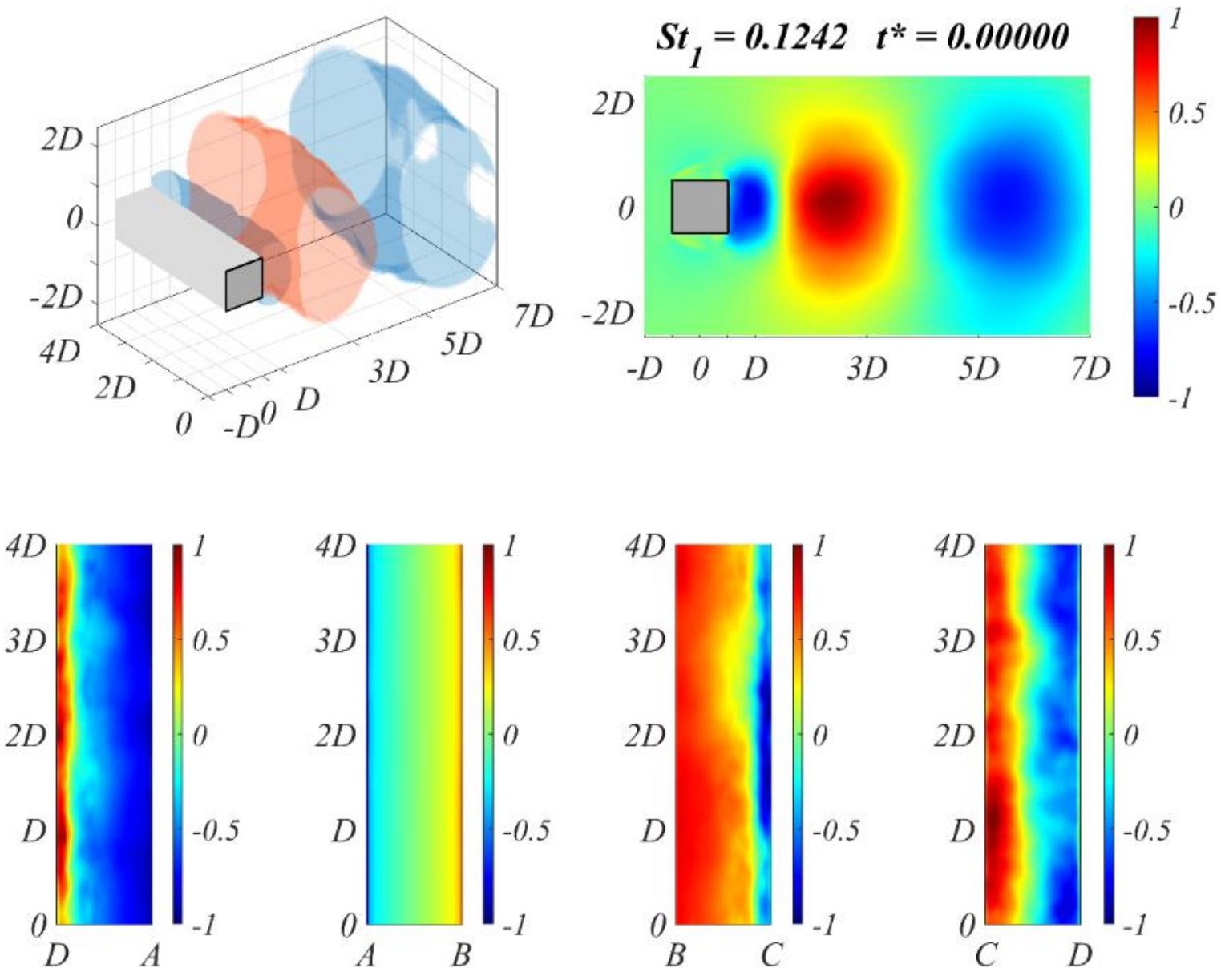

AI.2: Normalized mode shapes *(-1 to 1)* of $\boldsymbol{M_1}$ *($St_1$=0.1242)* of v and the prism walls: iso-surfaces ±*0.25* of *v* (top left); mid-prism-span slice of *v* (top right); the bottom (DA), upstream (AB), top (BC), and downstream (CD) walls, respectively (bottom from left to right). Multimedia file slowed by a factor of *500*.

## *AI.3* $M_1$ *- w*

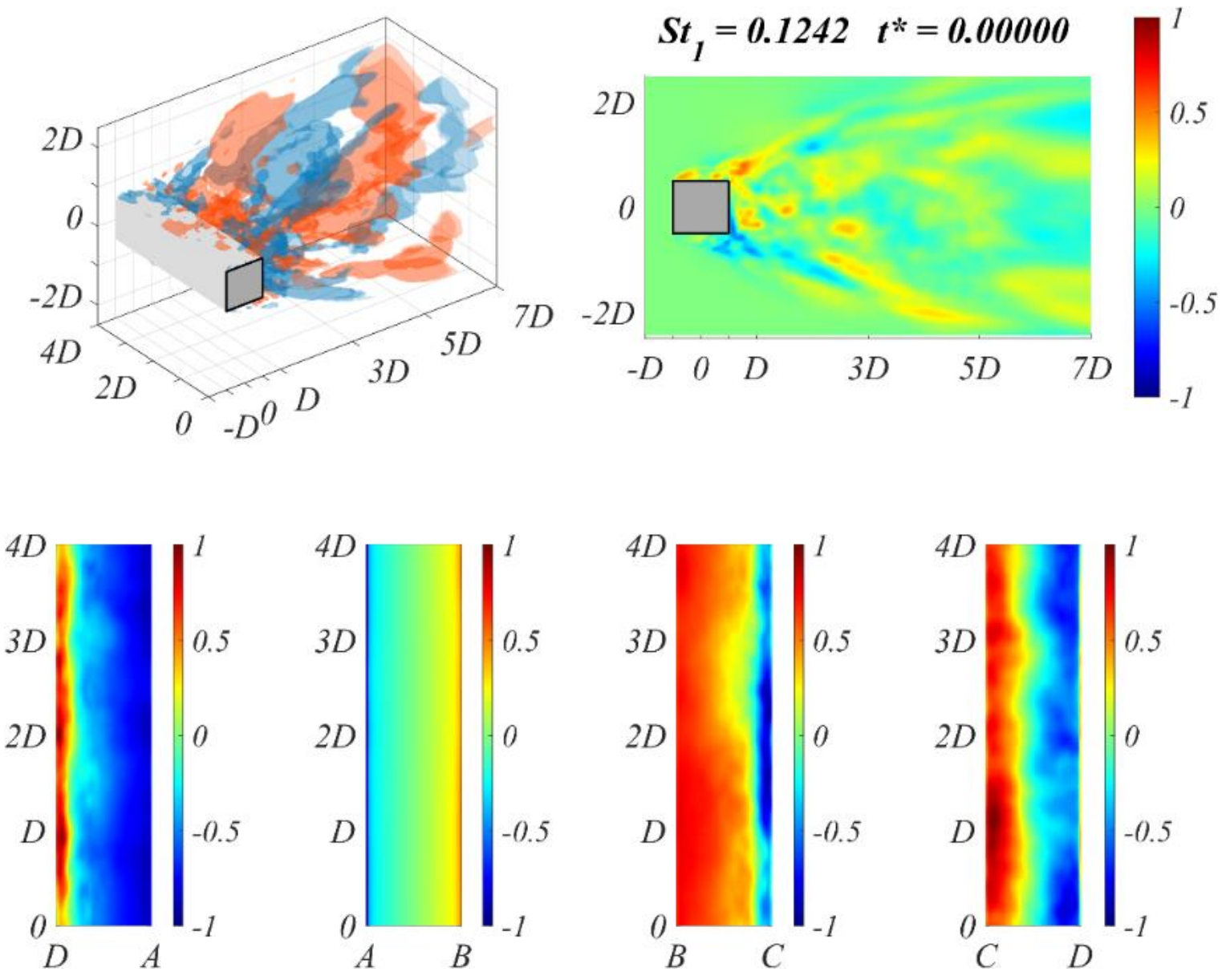

AI. 3: Normalized mode shapes *(-1 to 1)* of $\boldsymbol{M_1}$ *($St_1$=0.1242)* of *w* and the prism walls: iso-surfaces ±*0.25* of *w* (top left); mid-prism-span slice of *w* (top right); the bottom (DA), upstream (AB), top (BC), and downstream (CD) walls, respectively (bottom from left to right). Multimedia file slowed by a factor of *500*.

## *AI.4 $M_2$ - |U|*

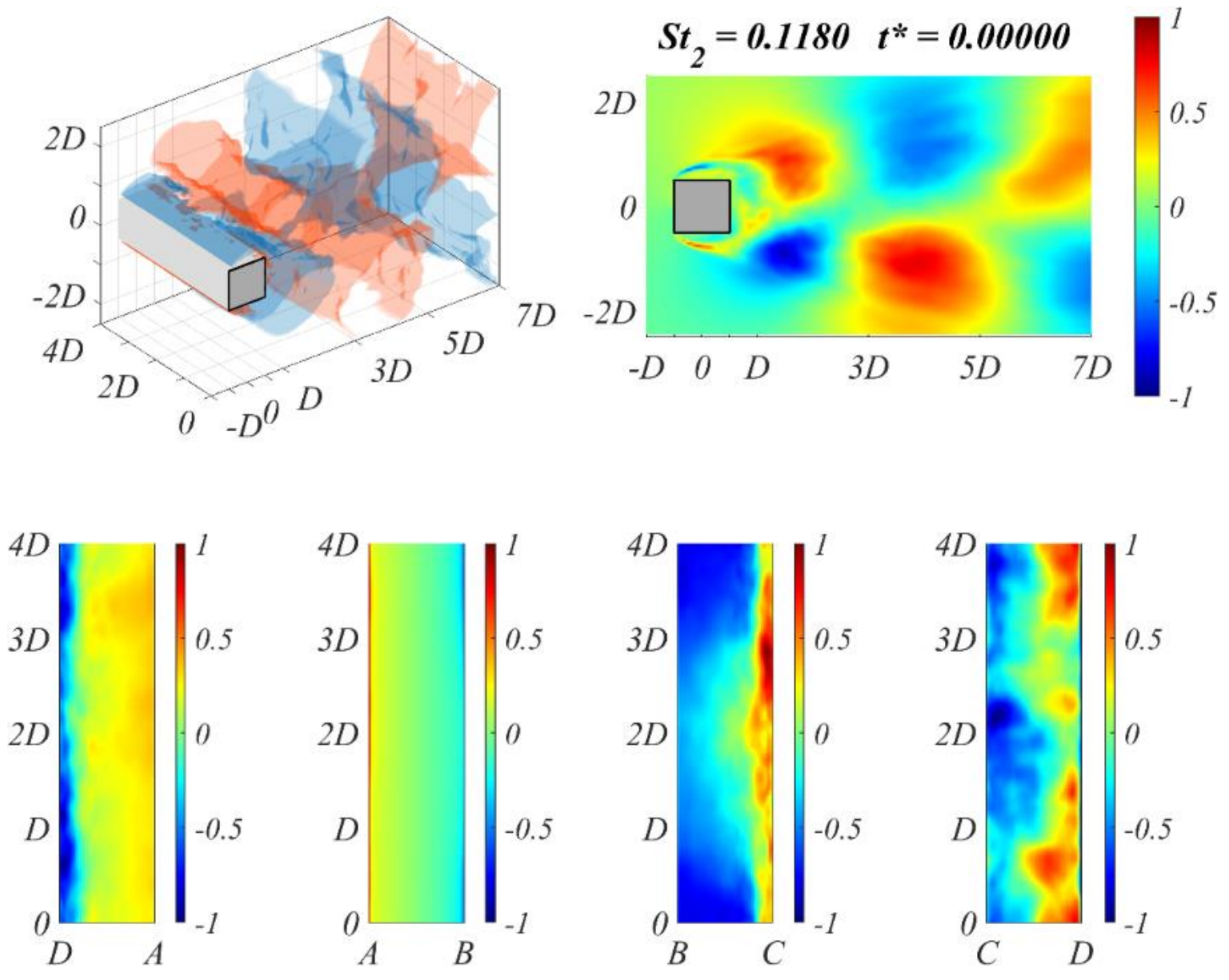

AI. 4: Normalized mode shapes *(-1 to 1)* of $\boldsymbol{M_2}$ *($St_2$=0.1180)* of |*U*| and the prism walls: iso-surfaces ±*0.25* of |*U*| (top left); mid-prism-span slice of |*U*| (top right); the bottom (DA), upstream (AB), top (BC), and downstream (CD) walls, respectively (bottom from left to right). Multimedia file slowed by a factor of *500*.

### *AI.5 $M_4$ - |U|*

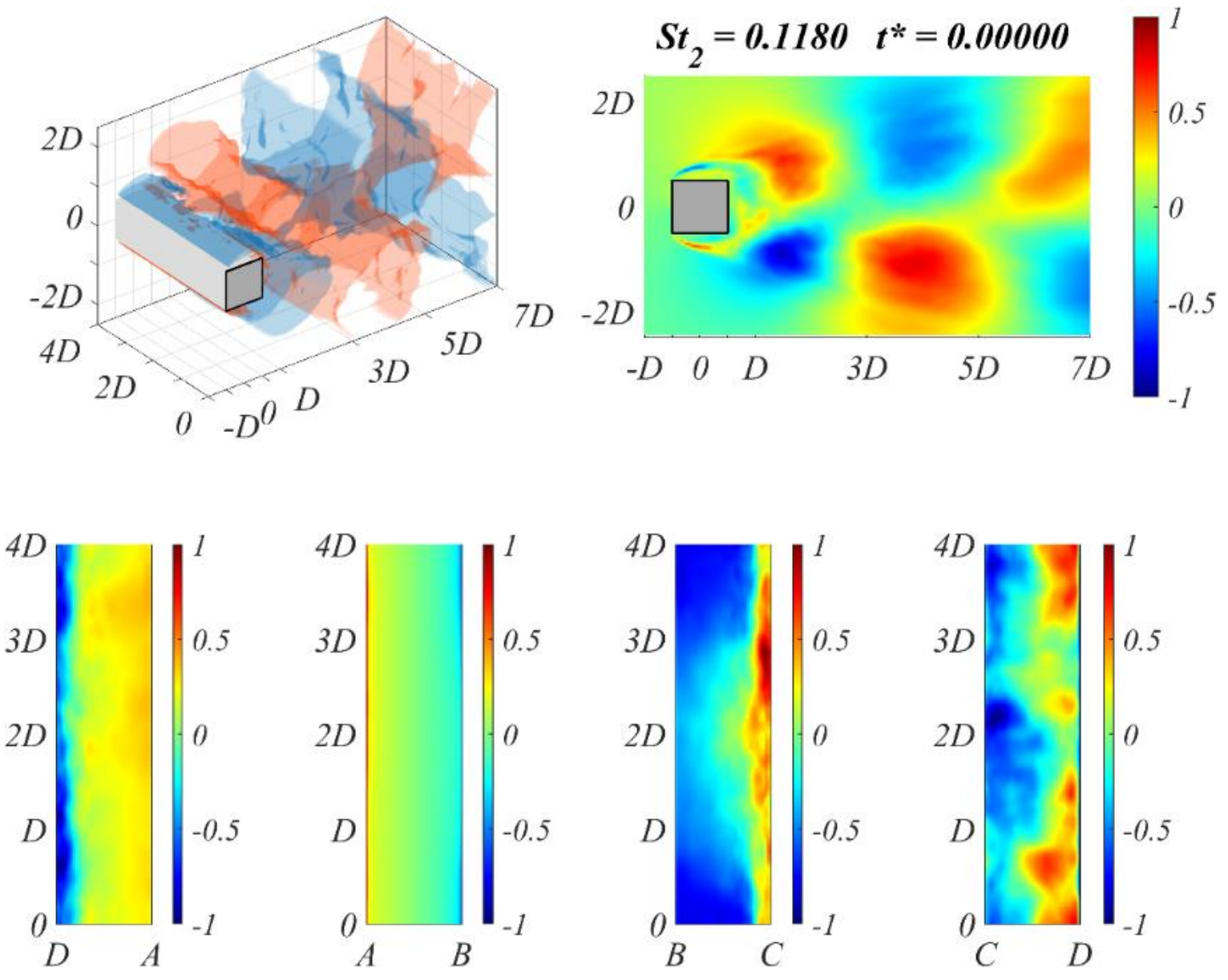

AI. 5: Normalized mode shapes *(-1 to 1)* of ***$M_4$*** *($St_4$=0.1304)* of |*U*| and the prism walls: iso-surfaces ±*0.25* of |*U*| (top left); mid-prism-span slice of |*U*| (top right); the bottom (DA), upstream (AB), top (BC), and downstream (CD) walls, respectively (bottom from left to right). Multimedia file slowed by a factor of *500*.

### *AI.6* $M_5$ - $\tilde{\Omega}_R$

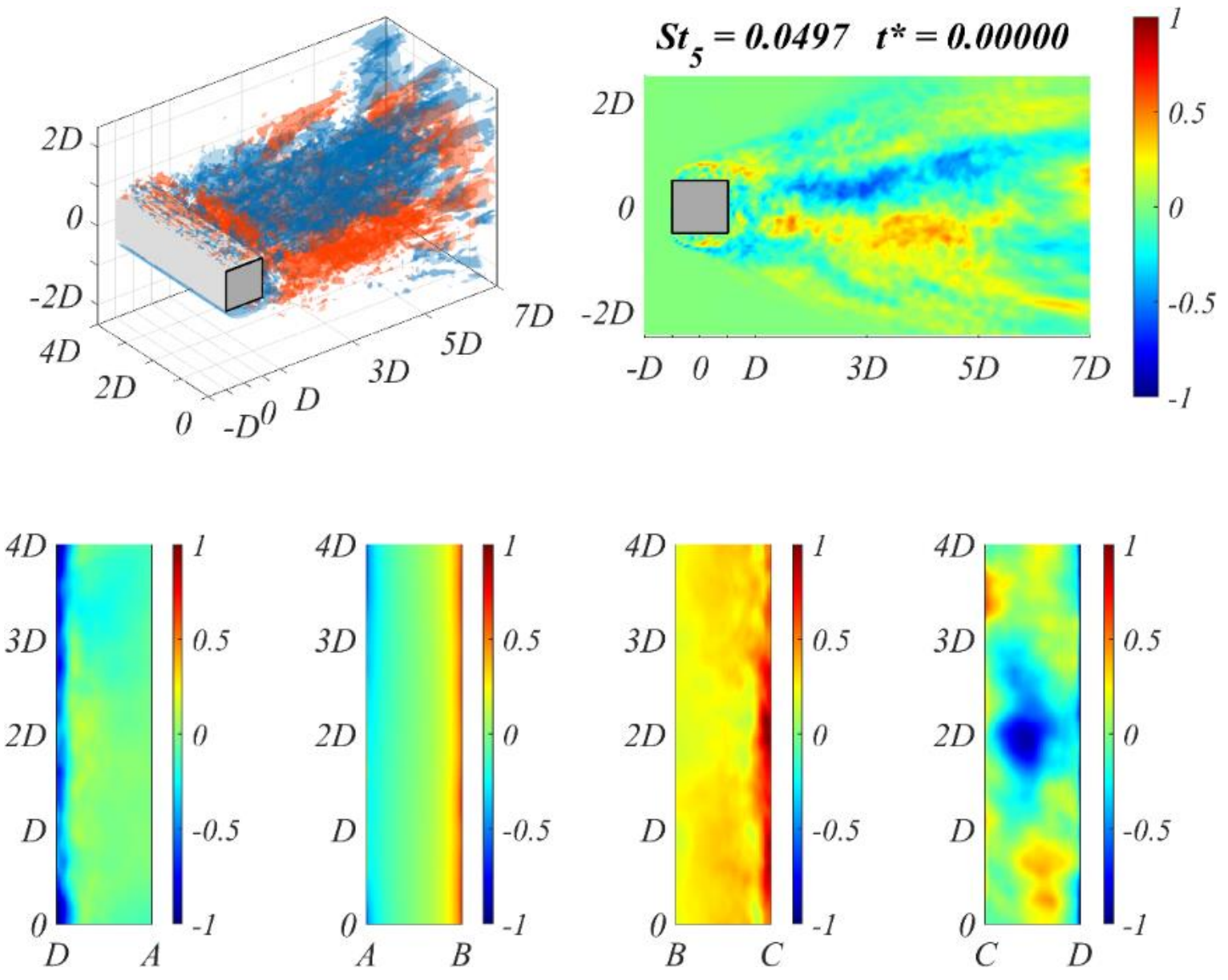

AI. 6: Normalized mode shapes *(-1 to 1)* of $\boldsymbol{M_5}$ *(*$St_5$*=0.0497)* of $\tilde{\Omega}_R$ and the prism walls: iso-surfaces ±*0.25* of $\tilde{\Omega}_R$ (top left); mid-prism-span slice of $\tilde{\Omega}_R$ (top right); the bottom (DA), upstream (AB), top (BC), and downstream (CD) walls, respectively (bottom from left to right). Multimedia file slowed by a factor of *500*.

## *AI.7 $M_7$ - P*

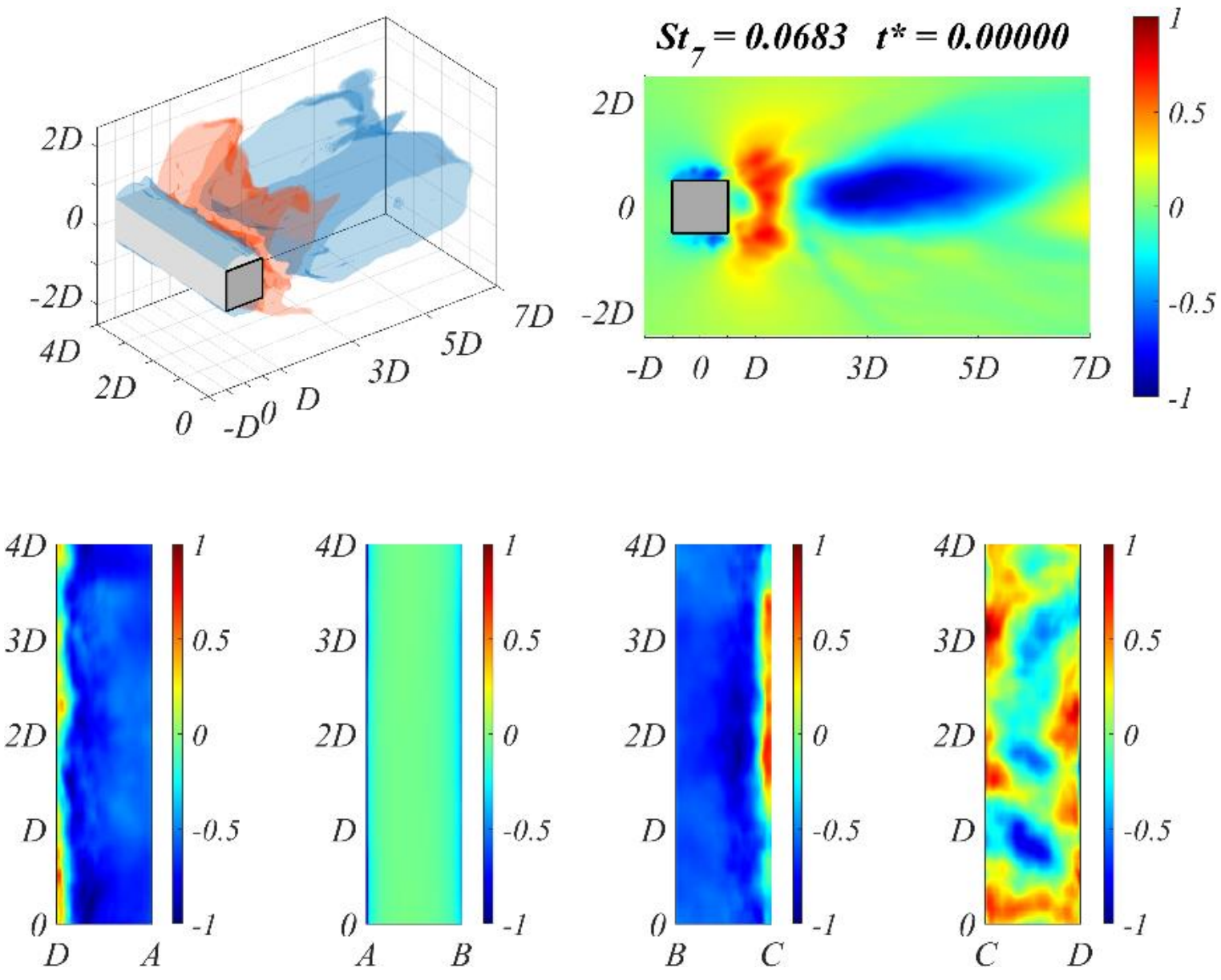

AI. 7: Normalized mode shapes *(-1 to 1)* of **$M_7$** *($St_7$=0.0683)* of P and the prism walls: iso-surfaces ±*0.25* of *P* (top left); mid-prism-span slice of *P* (top right); the bottom (DA), upstream (AB), top (BC), and downstream (CD) walls, respectively (bottom from left to right). Multimedia file slowed by a factor of *500*.

*AI.8 $M_6$ - |U|*

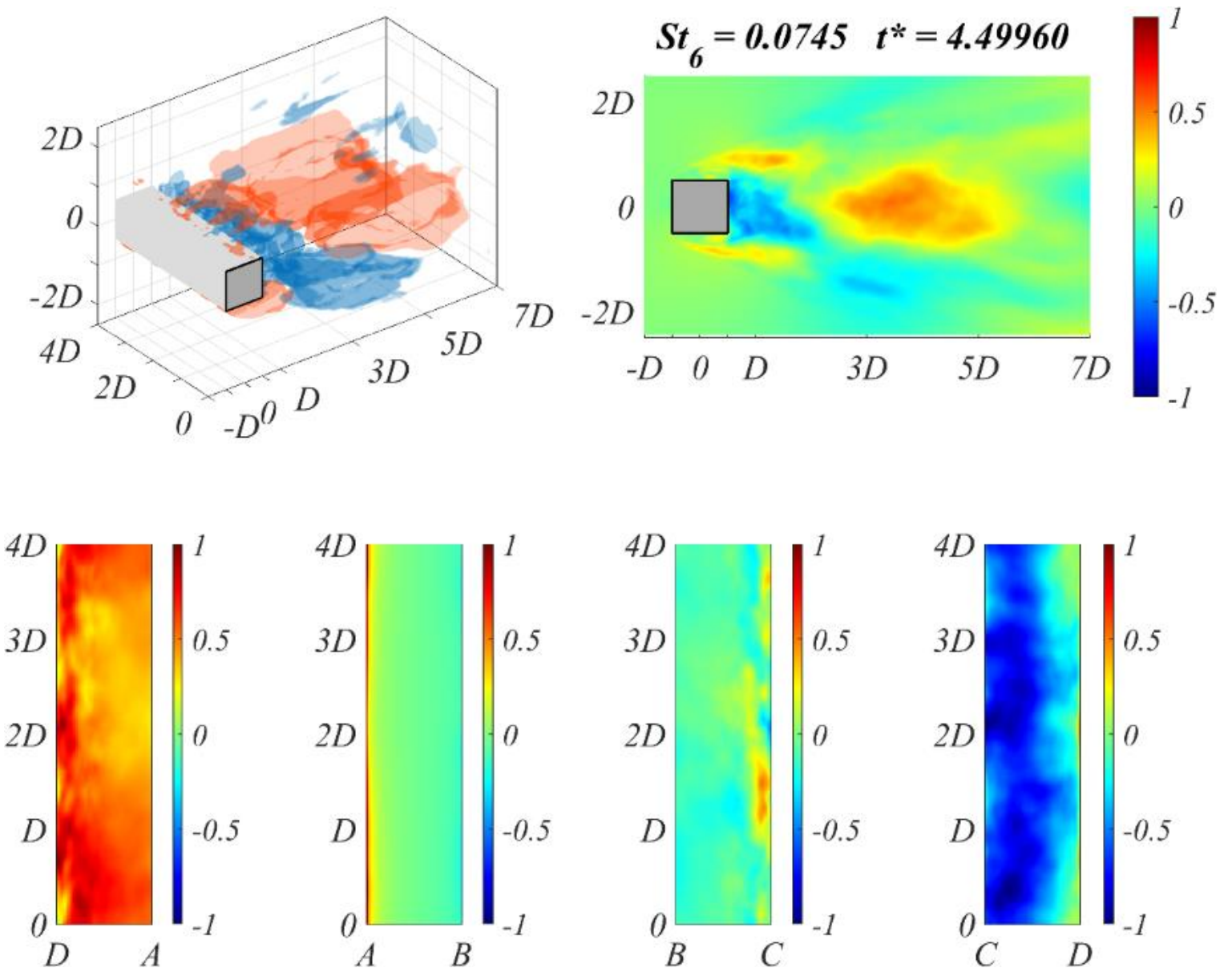

AI. 8: Normalized mode shapes *(-1 to 1)* of $\boldsymbol{M_6}$ *($St_6$=0.0745)* of |*U*| and the prism walls: iso-surfaces ±*0.25* of |*U*| (top left); mid-prism-span slice of |*U*| (top right); the bottom (DA), upstream (AB), top (BC), and downstream (CD) walls, respectively (bottom from left to right). Multimedia file slowed by a factor of *500.*

## *AI.9 $M_6$ - P*

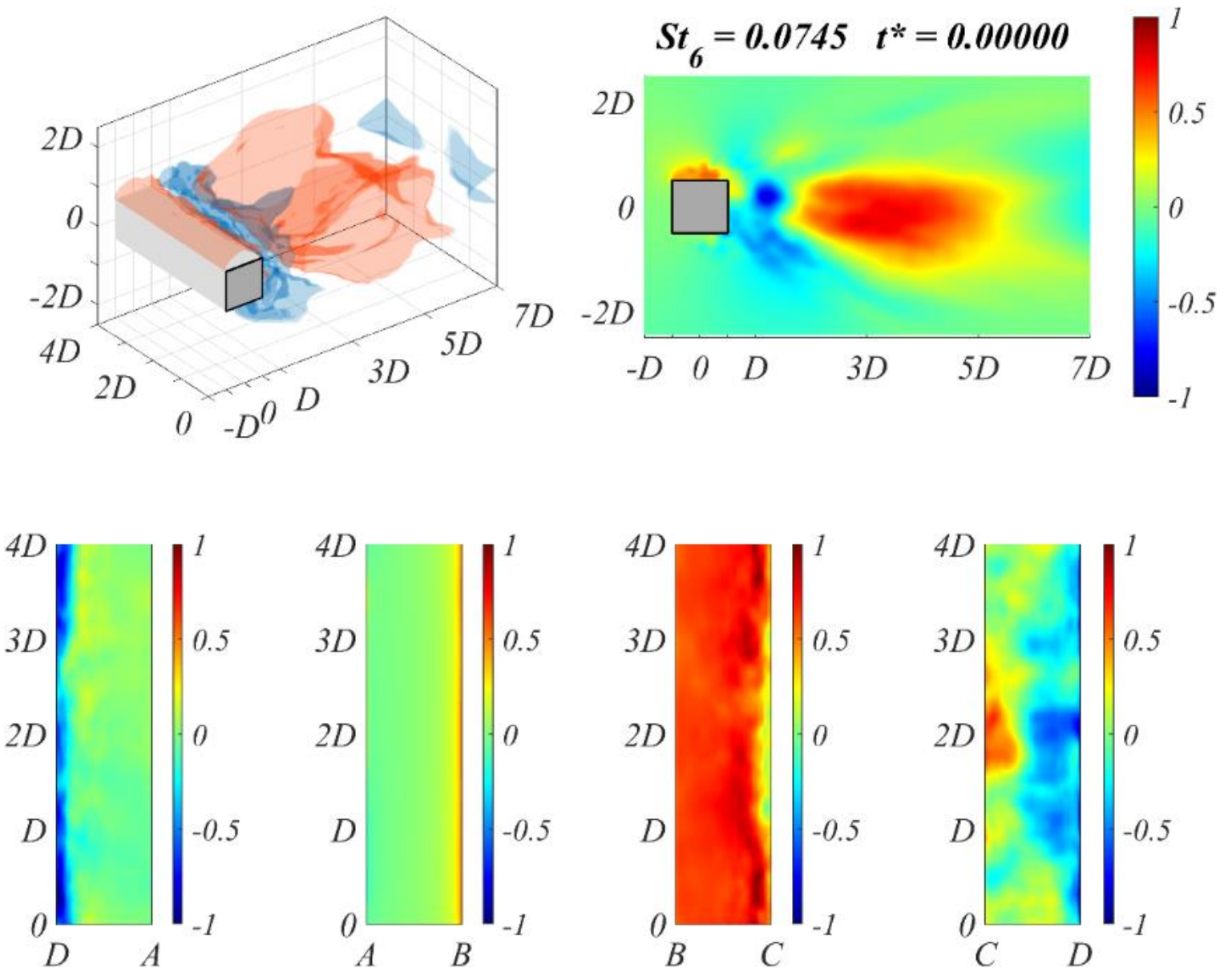

AI. 9: Normalized mode shapes *(-1 to 1)* of $\boldsymbol{M_6}$ *($St_6$=0.0745)* of *P* and the prism walls: iso-surfaces ±*0.25* of *P* (top left); mid-prism-span slice of *P* (top right); the bottom (DA), upstream (AB), top (BC), and downstream (CD) walls, respectively (bottom from left to right). Multimedia file slowed by a factor of *500*.

## *AI.10* $M_9$ - |*U*|

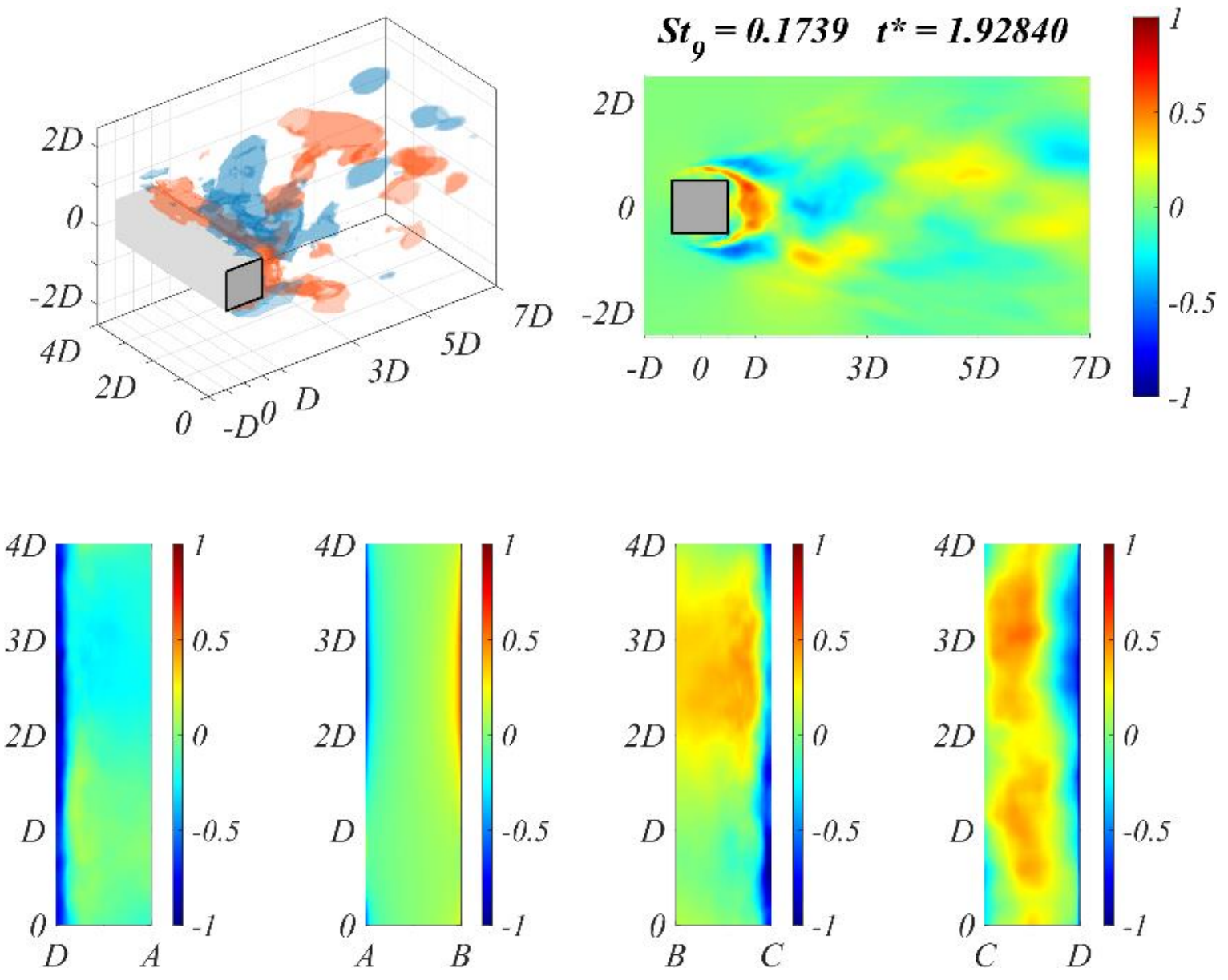

AI. 10: Normalized mode shapes *(-1 to 1)* of $\boldsymbol{M_9}$ *($St_9$=0.1739)* of |*U*| and the prism walls at: iso-surfaces ±*0.25* of |*U*| (top left); mid-prism-span slice of |*U*| (top right); the bottom (DA), upstream (AB), top (BC), and downstream (CD) walls, respectively (bottom from left to right). Multimedia file slowed by a factor of *500*.

## *AI.11 $M_{13}$* - |*U*|

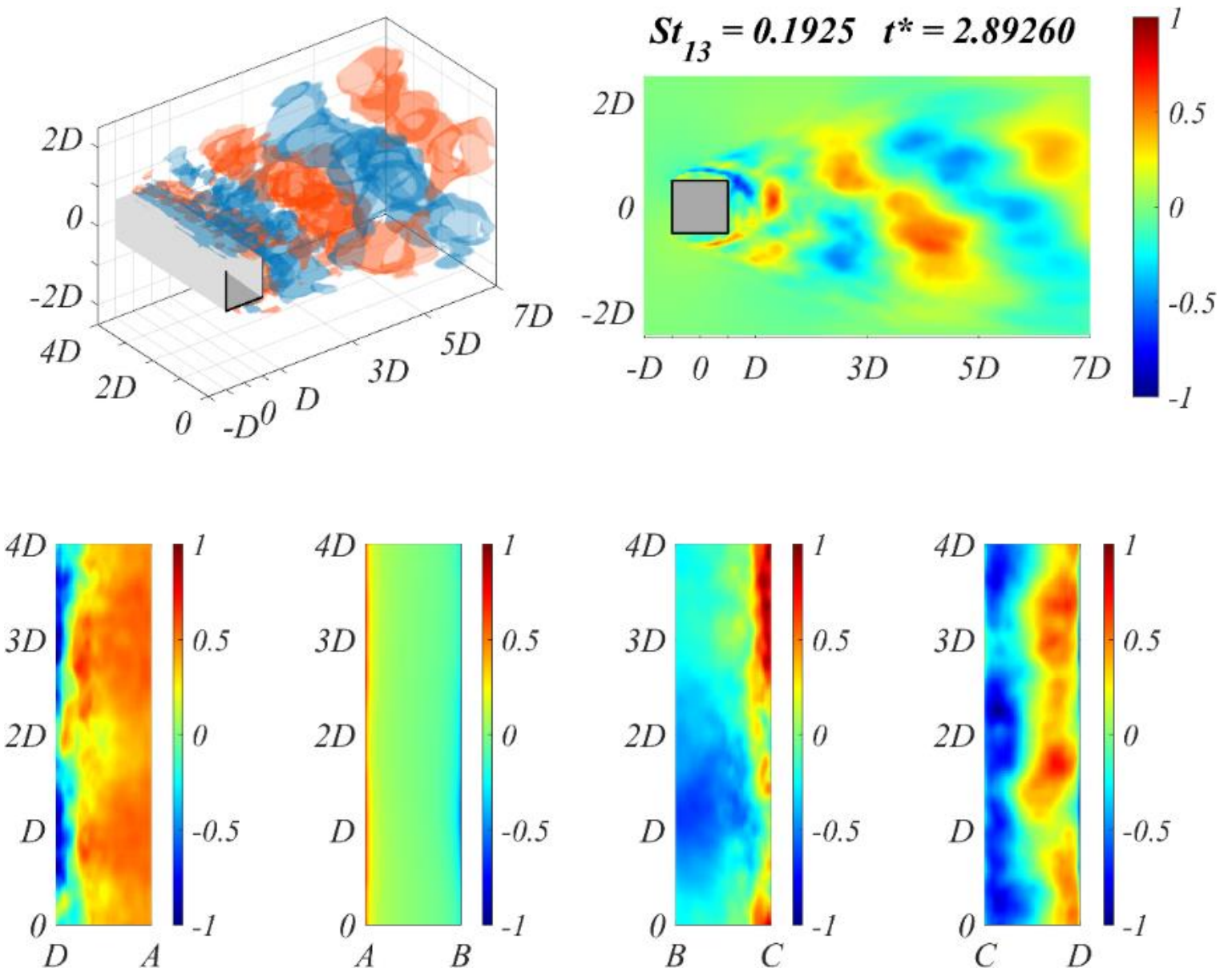

AI. 11: Normalized mode shapes *(-1 to 1)* of $\boldsymbol{M_{13}}$ *($St_{13}$=0.1925)* of |*U*| and the prism walls: iso-surfaces ±*0.25* of |*U*| (top left); mid-prism-span slice of |*U*| (top right); the bottom (DA), upstream (AB), top (BC), and downstream (CD) walls, respectively (bottom from left to right). Multimedia file slowed by a factor of *500*.

## Funding

The work described in this paper was supported by the Research Grants Council of the Hong Kong Special Administrative Region, China (Project No. 16207719), the Fundamental Research Funds for the Central Universities of China (Project No. 2021CDJQY-001), the National Natural Science Foundation of China (Project No. 51908090 and 42175180), the Natural Science Foundation of Chongqing, China (Project No. cstc2019jcyj-msxmX0565 and cstc2020jcyj-msxmX0921), the Key project of Technological Innovation and Application Development in Chongqing (Project No. cstc2019jscx-gksbX0017), and the Innovation Group Project of Southern Marine Science and Engineering Guangdong Laboratory (Project No. 311020001).

## Conflict of Interest

The authors declare that they have no conflict of interest.

## Availability of Data and Material

The datasets generated during and/or analyzed during the current work are restricted by provisions of the funding source but are available from the corresponding author on reasonable request.

## Code Availability

The custom code used during and/or analyzed during the current work are restricted by provisions of the funding source.

## Author Contributions

All authors contributed to the study conception and design. Funding, project management, and supervision were led by Tim K.T. Tse and Zengshun Chen and assisted by Xuelin Zhang. Material preparation, data collection, and formal analysis were led by Cruz Y. Li and Zengshun Chen and assisted by Asiri Umenga Weerasuriya, Yunfei Fu, and Xisheng Lin. The first draft of the manuscript was written by Cruz Y. Li and all authors commented on previous versions of the manuscript. All authors read, contributed, and approved the final manuscript.

## Compliance with Ethical Standards

All procedures performed in this work were in accordance with the ethical standards of the institutional and/or national research committee and with the 1964 Helsinki declaration and its later amendments or comparable ethical standards.

## Consent to Participate

Informed consent was obtained from all individual participants included in the study.

## Consent for Publication

Publication consent was obtained from all individual participants included in the study.